\documentclass[fleqn,10pt]{wlscirep}
\usepackage[utf8]{inputenc}
\usepackage[T1]{fontenc}

\usepackage{tabulary}
\usepackage{listings}
\usepackage{subfigure}

\definecolor{LightGray}{rgb}{0.97,0.97,0.97}
\definecolor{MidnightBlue}{rgb}{0.1, 0.1, 0.44}
\definecolor{DarkOliveGreen}{rgb}{0.33, 0.42, 0.18}
\definecolor{Sepia}{rgb}{0.44, 0.26, 0.08}
\definecolor{Purple}{rgb}{0.5, 0.0, 0.5}
\lstdefinelanguage{SPARQL}{
	basicstyle=\small\ttfamily,
	backgroundcolor=\color{LightGray},
	columns=fullflexible,
	breaklines=true,
	sensitive=true,
	frame=bt,
	aboveskip=1em,
	belowskip=1em,
	xleftmargin=.5em,
	xrightmargin=.5em,
	framexleftmargin=.5em,
	framextopmargin=.5em,
	framexbottommargin=.5em,
	framexrightmargin=.5em,
	tabsize = 2,
	showstringspaces=false,
	morecomment=[n][\color{blue}]{<http}{>}, 
	morestring=[b][\color{DarkOliveGreen}]{\"}, 
	keywordsprefix=?,
	classoffset=0,
	keywordstyle=\color{Sepia},
	morekeywords={},
	classoffset=1,
	keywordstyle=\color{Purple},
	morekeywords={rdf,rdfs,owl,xsd,purl,cso,cpo,ffo,aami,hasProperty,hasSymbolData, ala},
	classoffset=2,
	keywordstyle=\color{MidnightBlue},
	morekeywords={
	SELECT,CONSTRUCT,DESCRIBE,ASK,WHERE,FROM,NAMED,PREFIX,BASE,OPTIONAL,
	FILTER,GRAPH,LIMIT,OFFSET,SERVICE,UNION,EXISTS,NOT,BINDINGS,MINUS,a
	,curl, -o,@prefix}
	}

\lstdefinelanguage{XML}
{
   basicstyle=\small\ttfamily,
   backgroundcolor=\color{LightGray},
   columns=fullflexible,
   sensitive=true,
   morestring=[b][\color{DarkOliveGreen}]{\"}, 
   moredelim=[s][\bfseries\color{maroon}]{<}{\ },
   moredelim=[s][\bfseries\color{maroon}]{</}{>},
   moredelim=[l][\bfseries\color{maroon}]{/>},
   moredelim=[l][\bfseries\color{maroon}]{>},
   morecomment=[l][\color{gray}]{\#}, 
	morecomment=[n][\color{blue}]{<http}{>}, 
	morestring=[b][\color{DarkOliveGreen}]{\"}, 
   commentstyle=\color{darkolivegreen},
   stringstyle=\color{blue},
   identifierstyle=\color{red},
   numbers=left,
   tabsize=2,
   frame=bt,
   rulecolor=\color{black},
   breaklines=true,
   captionpos=b
}

\title{Better force fields start with better data - \\A data set of cation dipeptide interactions}

\author[1,*]{Xiaojuan Hu}
\author[1]{Maja-Olivia Lenz-Himmer}
\author[1,*]{Carsten Baldauf}
\affil[1]{Fritz-Haber-Institut der Max-Planck-Gesellschaft, Faradayweg 4-6, 14195 Berlin, Germany}

\affil[*]{corresponding authors: Xiaojuan Hu (xhu@fhi.mpg.de) and Carsten Baldauf (baldauf@fhi.mpg.de)}

\begin{abstract}
\noindent
We present a data set from a first-principles study of amino-methylated and acetylated (capped) dipeptides of the 20 proteinogenic amino acids -- including alternative possible side chain protonation states and their interactions with selected divalent cations (Ca$^{2+}$, Mg$^{2+}$ and Ba$^{2+}$).
The data covers 21,909 stationary points on the respective potential-energy surfaces in a wide relative energy range of up to 4 eV (390 kJ/mol). 
Relevant properties of interest, like partial charges, were derived for the conformers.
The motivation was to provide a solid data basis for force field parameterization and further applications like machine learning or benchmarking.
In particular the process of creating all this data on the same first-principles footing, i.e. density-functional theory calculations employing the generalized gradient approximation with a van der Waals correction, makes this data suitable for data-driven force field development. 
To make the data accessible across domain borders and to machines, we formalized the metadata in an ontology.

\end{abstract}
\begin{document}

\flushbottom
\maketitle

\thispagestyle{empty}


\section*{Background \& Summary}


Metal cations are essential to life: one third of the proteins in the human body require metal cofactors.\cite{permyakov2009metalloproteomics, bertini2007biological}
By shaping the structure of proteins, cations affect biological processes like molecular recognition or enzyme activity.
Understanding the structure, dynamics, and function of metalloproteins is in the ongoing focus of many researchers, we summarize a few examples that involve simulation approaches:
Tamames \textit{et al.} analyzed zinc coordination spheres in a data set from the Protein Data Bank and complemented with DFT-B3LYP calculations.\cite{tamames2007analysis}
Sala \textit{et al.} investigated folding of Pyrococcus furiosus rubredoxin (PfRd), which includes an iron ion, with classical molecular dynamics (MD) simulations.\cite{sala2018molecular}
A calcium binding site in the blood protein von Willebrand Factor (VWF) regulates force-triggered unfolding for cleavage and therewith its activity in primary hemostasis, as illustrated by classical force-probe MD simulations.\cite{zhou2011novel}
Gogoi \textit{et al.} investigated protein-metal ion binding affinities by analysing MD simulations of 49 different cation-proteins complexes.\cite{gogoi2016heterogeneous}
Metal cations can alter peptide structure by interacting with backbones and thereby enforcing non-Ramachandran geometries. \cite{baldauf2013cations} 
Cations can, by repulsion or attraction, also substantially reduce the conformational flexibility of functional sidechains.\cite{de2017mapping, ropo2016trends}

MD simulations of biomolecules typically rely on additive force fields, where distinct terms describe bonded and non-bonded interactions based on empirically derived parameters.
Studies have shown that the accuracy of force fields is especially limited when describing interactions involving ionic species.\cite{vitalini2015dynamic, schneider2018relative, maksimov2021conformational, marianski2016assessing} 
In particular non-bonded interactions are critical, but of course the effect that nearby located cations exert on bonds is almost impossible to grasp by the combination of bonded and non-bonded interactions in a general-purpose force field. 
Modeling of electrostatic interactions via pairwise Coulomb potentials is based on assigning partial charges to atoms.\cite{wang2001automatic}
Partial charges are derived by: 
(i)~fitting to experimental data, e.g. by fitting partial charges to reproduce hydration free enthalpies\cite{oostenbrink2004biomolecular} (GROMOS and OPLS prior 2005) and 
(ii)~deriving partial charges from QM calculations (Amber and Charmm).\cite{wang2000well,riniker2018fixed}

\begin{figure}[ht]
\vspace{0.0in}
\begin{center}
\includegraphics[width=0.8\textwidth]{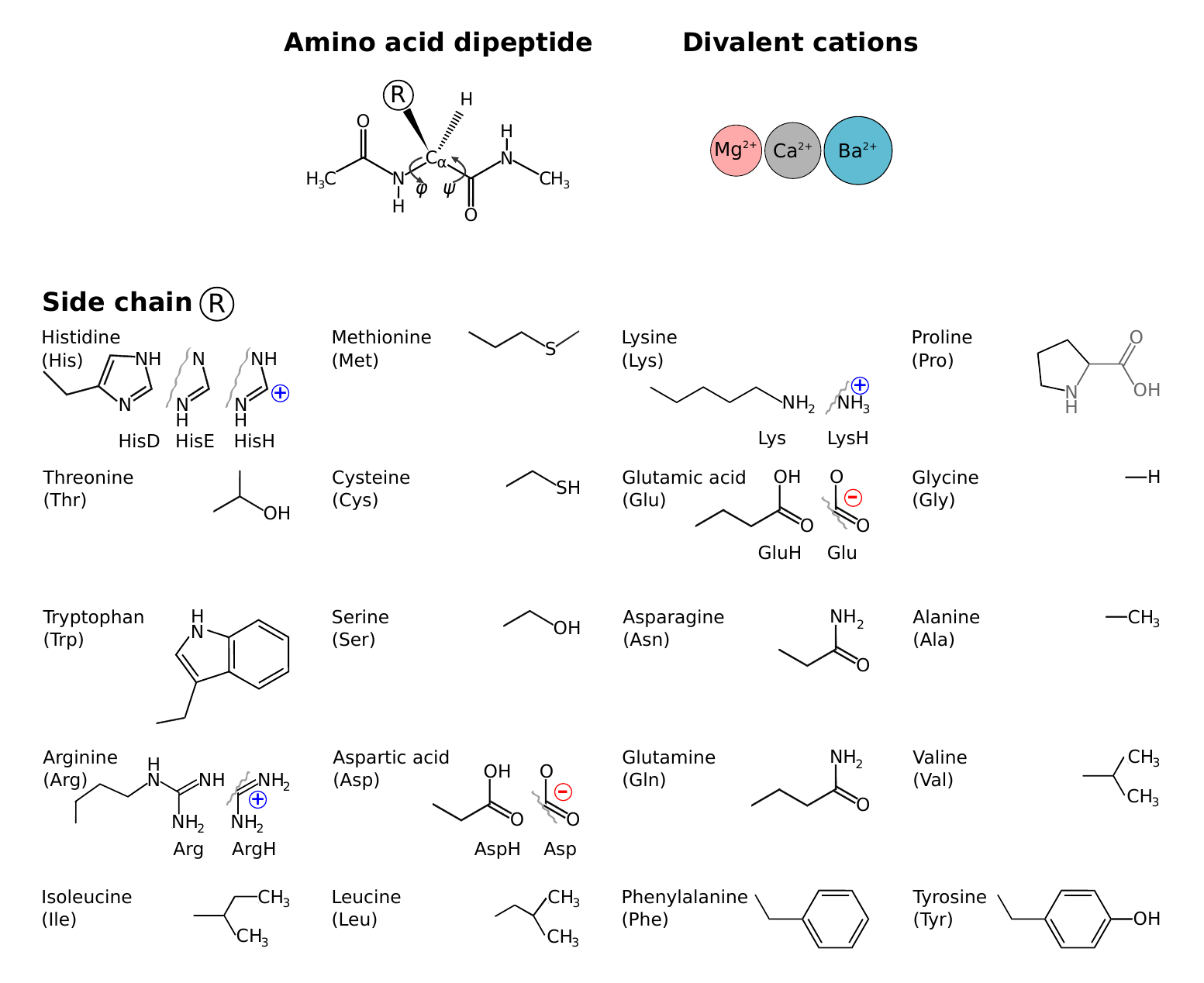}
\end{center}
\vspace{-0.2in}
\setlength{\belowcaptionskip}{0.2in}
\caption{The molecular systems in this study are dipeptides of the 20 proteinogenic amino acids that (with the exception of Pro) differ in the side chain \textbf{R}. Where applicable, different protonation states were considered.}
\label{Molecular_systems}
\vspace{-0.2in}
\end{figure}

The reliability of a force field also depends on the physics behind the formulation. 
The failures of established biomolecular force fields when describing cation-peptide systems may result from a central underlying assumption --  modeling atoms by fixed point charges and neglecting charge transfer and polarization effects, while both are crucial to ionic systems.\cite{allen2004energetics, roca2003theoretical, zeng2013f130l, li2011structure} 
Introducing more physics to the model appears a promising route to improve force fields: The inclusion of electronic polarization and charge transfer plays a central role in the next generations of biomolecular force fields.\cite{xie2009coupled,ngo2015quantum,amin2020benchmarking} 
However, including additional terms leads to force fields with way more parameters, which makes parameterization more challenging, \cite{liang1996parameter, faller1999automatic} in particular in the absence of high-resolution experimental data of less stable conformations, i.e. higher-energy structures. \cite{cisneros2014classical}
Summarizing, we see three main challenges:
\begin{itemize}
    \item The availability of sufficiently-accurate electronic-structure data as well as choosing the ``right ways'' to derive e.g. partial charges from it.
    \item Designing the formulation of next-generation force fields that also include, for example, charge transfer and polarization. 
    \item Finding sets of parameters (force fields) for such potentials in the absence of experimental data at sufficient spatial and time resolution.
\end{itemize}
Comprehensive computational data at an appropriately accurate level of theory has the potential to substantially increase the predictive power of force fields. \cite{smith2017ani}
Thorough studies have deepened our understanding of the conformational basics of individual building blocks, e.g. \cite{rezac2018toward, jurevcka2006benchmark, goerigk2017look, dohm2018comprehensive, yu2009extensive, kishor2008structural, selvarengan2004potential, csaszar1999ab, schlund2008conformational, riffet2011acid, baek2011density, floris2012density}. 
However, these studies are highly diverse with regards to the approximations made to model and to search the potential energy surfaces (PES) of the respective molecular systems; furthermore, the data is often not available. 
In order to provide amino acid data sets for force field development on consistent computational footing, we extend previous work\cite{ropo2016first} by focusing on dipeptides as models of amino acid building blocks in polypeptide chains in complex with the divalent cations Mg$^{2+}$, Ca$^{2+}$, and Ba$^{2+}$, which play prominent roles in physiology:

\begin{itemize}
\item Mg$^{2+}$ takes structural, catalytic, and regulatory roles\cite{huang1995biostructural} regulating ion channels, mitochondrial function, and cell's pH and volume.\cite{romani2011cellular}

\item Ca$^{2+}$ levels regulate muscle contraction, hormone secretion, metabolism, ion transport, division \textit{etc. }\cite{forsen1994calcium} Mg$^{2+}$ and Ca$^{2+}$ may compete for the same binding sites.\cite{grauffel2019cellular}

\item Ba$^{2+}$ can cause cardiac irregularities and affect the nervous system presumably by blocking potassium channels.\cite{mahmoud2012functionalized}
\end{itemize}
The data set we describe covers a wide range of molecular systems, see Figure~\ref{Molecular_systems}.
For the 21,909 stationary points, properties relevant to force field development were computed, details can be found in the Methods section.
Making the data FAIR\cite{wilkinson2016fair,wittenburg2020fair} -- as in findable, accessible, interoperable, and reusable -- is a challenge. 
In particular as we want to make the data available also to experts from other domains of science or to autonomous agents.
To that end, we make the data freely available and also providing ontologies.
An ontology defines a common vocabulary for researchers who need to share information by including machine-interpretable definitions of basic concepts in a domain and relations among them.\cite{noy2001ontology}
A common semantic vocabulary enables interoperability between resources and databases as well as data interpretation across data collections.
By the developed ontological representation of the data set, it can be connected to upper level concepts and thereby made machine-usable, which in turn enables automatic access and querying of the data.

\section*{Methods}






 

Figure~\ref{Molecular_systems} summarizes the molecular systems in this study. 
Including the protonation states, we have to consider 26 dipeptides in 4 complexation states (bare, Ca$^{2+}$, Mg$^{2+}$, Ba$^{2+}$) which results in the 104 systems for which our structure searches identified 21,909 stationary points.
For each of these stationary points, not only structure and energy are provided, but also further properties relevant to force field development, namely: 
van der Waals energies, interaction energies as well as electron densities and derived properties like the electrostatic potential, diverse partial charge models, and effective atomic volumes.
By that, our dipeptide-cation data set allows one to explicitly assess subtle, but important, effects of local changes in the electrostatic environment due to peptide-cation interaction.

\subsection*{Sampling method}
A hierarchical structure search that is described in detail in reference\cite{ropo2016first} was employed to locate stationary points on the the potential energy surfaces of the 104 molecular systems.  
The initial global conformational searches of all dipeptides with/without Ca$^{2+}$ were performed by a basin hopping search strategy\cite{wales1997global, wales1999global} using the OPLS-AA force field.\cite{jorgensen1996development}
Secondly, a refinement using density-functional theory calculations was performed.
All electronic-structure calculations were performed with the all-electron, full-potential code FHI-aims utilizing numeric atom-centered basis functions.\cite{blum2009ab, havu2009efficient, ren2012resolution}
The PBE generalized-gradient exchange-correlation functional\cite{perdew1996generalized} augmented by Tkatchenko's and Scheffler's pairwise van der Waals correction\cite{tkatchenko2009accurate} was employed, and is referred to as PBE+vdW throughout this work.
Stationary points that resulted from the FF-based pre-sampling were subjected to DFT-PBE+vdW relaxations with \texttt{light} settings. 
Next, a local first-principles based sampling step by \textit{ab initio} replica-exchange molecular dynamics (REMD)\cite{swendsen1986replica, sugita1999replica} employing DFT-PBE+vdW with \texttt{light} settings, was applied to the identified set of structures. 
Conformers were extracted every 10 steps from REMD trajectories and clustered with a $k$-means clustering algorithm.\cite{wong1979algorithm}
Obtained conformers went through relaxation with PBE+vdW (\texttt{light} computational settings), clustering and further relaxation with PBE+vdW (\texttt{tight} computational settings) to obtain the final conformational hierarchies. 
Initial structures of Mg$^{2+}$ and Ba$^{2+}$ binding dipeptides were obtained by substituting Ca$^{2+}$ cation in dipeptide binding a Ca$^{2+}$ cation.
Subsequently, those were put into the procedure from \textit{ab initio} REMD simulations to relaxation with PBE+vdW (\texttt{light} computational settings) to obtain final conformers as described before.
These structures were further relaxed by PBE+vdW with \texttt{tight} computational settings. 

\subsection*{Property calculations}
\begin{figure}[ht]
\vspace{0.0in}
\begin{center}
\includegraphics[width=0.7\textwidth]{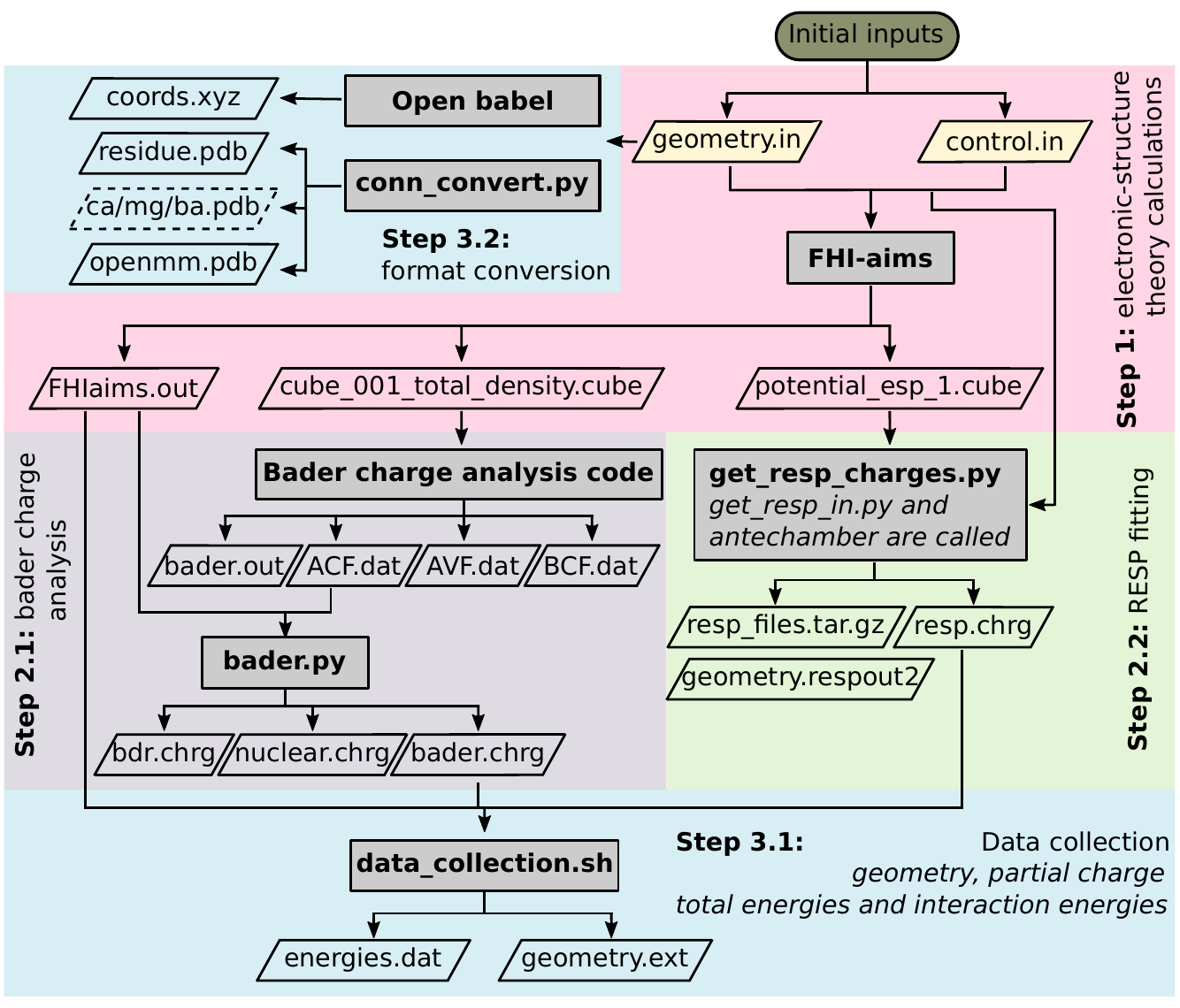}
\end{center}
\vspace{-0.2in}
\setlength{\belowcaptionskip}{0.2in}
\caption{Schematic representation of the workflow employed to derive properties of each conformers. Calculation steps were displayed in boxes with different background colour. Gray boxes indicate tools employed in each step. Parallelograms represent input and output files in each step. Links to custom codes are listed in Section~\textit{Code availability}. }
\label{Workflow}
\vspace{-0.2in}
\end{figure}

Property calculations were performed on all structures obtained by the sampling method described above. 
This includes also high energy conformers.
Figure~\ref{Workflow} shows the processes involved in the property calculations; the individual steps are described in detail below.
From the PBE+vdW DFT calculations with \texttt{tight} computational settings using FHI-aims, we collect in \textbf{Step~1} total energies, vdW energies, interaction energies, electron densities, electrostatic potential, Hirshfeld partial charges,\cite{hirshfeld1977bonded} and effective atomic volumes. 

Based on the effective atomic volumes $V^\mathrm{eff}$ per atom we provide, the \underline{effective vdW radii} and the \underline{polarizability} of an atom in a molecule can be calculated as follows:\cite{tkatchenko2009accurate, distasio2014many}
\begin{align}
R^0_\mathrm{eff} =& R^0_\mathrm{free} \left (\frac{V^\mathrm{eff}}{V^\mathrm{free}} \right )^{1/3} \\
\alpha^0_\mathrm{eff} =& \alpha^0_\mathrm{free} \left (\frac{V^\mathrm{eff}}{V^\mathrm{free}} \right ) \\
\frac{V^{\mathrm{eff}}_{i}}{V^{\mathrm{free}}_{i}} = & \frac{\int r^3\omega_i(\vec{r})n(\vec{r})d^3\vec{r}}{\int r^3n^{\mathrm{free}}_{i}(\vec{r})d^3\vec{r}}
\end{align}
\noindent
in which, $R^0_\mathrm{free}$ and $\alpha^0_\mathrm{free}$ are the vdW radii of reference free-atom and static dipole polarizability (which can be taken from either experimental data or high-level quantum chemical calculations), respectively. 
$V^\mathrm{free}$ is the volume of the free atom \textit{in vacuo}, $r^3$ is the cube of the distance from the nucleus of atom $i$, $\omega_i(\vec{r})$ is the Hirshfeld atomic partitioning weight for atom $i$, $n(\vec{r})$ is the total electron density, and $n^{\mathrm{free}}_{i}(\vec{r})$ is the electron density of the free atom $i$.


The basic property resulting from a DFT calculation is the \underline{electron density}, which -- for each entry in our data set -- was stored on a discrete grid of points with a spacing of 0.05 {\AA} in a rectangular volume, which spans the whole molecule plus 14 Bohr (7.4\,\AA) beyond the outermost nuclei. 
The \underline{electrostatic potential} exerted by a molecule on its environment may be used to derive partial charges. 
To that end, for each entry in the data set, five molecular surfaces were created by increasing the van der Waals radii of all atoms in the molecule (molecule with cation) by factors between 1.4 and 2.0. 
Points on these surfaces were represented in a cubic grid of each 35 grid points in $x$, $y$, and $z$ direction. 
For these points, the electrostatic potential was evaluated.
For biomolecular force fields, atomic partial charges are a crucial ingredient for computing the pairwise Coulomb term of the non-bonded interactions.
We provide three types of partial charges: 

\begin{itemize}
    \item 
\underline{Hirshfeld atomic charges}, computed by FHI-aims, were derived based on the Hirshfeld partitioning scheme.\cite{tkatchenko2009accurate, hirshfeld1977bonded}
The Hirshfeld atomic charge $q_i$ of atom $i$ is given by
\begin{equation}
q_i = Z_i - \int n_i(\vec{r})d^3\vec{r}
\end{equation}

where $Z_i$ refers to the corresponding atomic number, and $n_i(\vec{r})$ is the associated electron density associated with atom $i$.
\begin{equation}
n_i(\vec{r}) = \omega_i(\vec{r})n(\vec{r})
\end{equation}
where $n(\vec{r})$ denotes the total electron density, $\omega_i(\vec{r})$ is the Hirshfeld atomic partitioning weight for atom $i$. $\omega_i(\vec{r})$ is given by
\begin{equation}
\omega_i(\vec{r}) = \frac{n^{\mathrm{free}}_{i}(\vec{r})}{\sum_{A}^{All atoms}n^{\mathrm{free}}_{A}(\vec{r}) }
\end{equation}

\item 
\underline{Bader charges} were being computed in \textbf{Step 2.1} using the Bader Charge Analysis tools\cite{henkelman2006fast, sanville2007improved, yu2011accurate} provided by the Henkelman group based on the electron density cube file produced in Step 1.
The atoms in molecules (AIM) partitioning method uses what is called zero flux surfaces to distribute electron density among the atoms. 
Such zero flux surface is a 2-D surface on which the charge density is a minimum perpendicular to the surface.
In molecular systems, the charge density typically reaches a minimum somewhere between pairs of neighboring nuclei.
This can be seen as the natural place to separate atoms from each other. 
These borders between atoms define the electron density region associated with a given atom, from which the partial charges are being calculated. 

\item
 In \textbf{Step 2.2}, \underline{RESP partial charges}\cite{bayly1993well, singh1984approach, fox1998application} were computed using Antechamber\cite{wang2001antechamber} from the AmberTools package.\cite{salomon2013overview} A two-stage restrained electrostatic potential (RESP) fitting procedure \cite{bayly1993well} was employed as implemented in Antechamber.  
\end{itemize}
In the final \textbf{Steps 3.1 and 3.2}, data was collected and files converted to established formats.
Geometry information is provided in three formats: 
the FHI-aims input format,  
the xyz format generated by Open\,Babel,\cite{o2011open} and
PDB files that are readable by the CHARMM-GUI portal\cite{jo2008charmm} and the openMM7 package.\cite{eastman2017openmm}
Connectivity and atom type information -- needed for the PDB format -- was gathered based on atomic distances by the Python script \texttt{conn\_convert.py}.
Furthermore, energies and partial charges were tabulated for convenient usage.
Interaction energies E$_{\mathrm{inter}}$ were calculated as follows:
\begin{equation}
E_{\mathrm{inter}} =  E_{\mathrm{complex}} - E_{\mathrm{dipeptide}} - E_{\mathrm{cation}}
\end{equation}
where 
$E_{\mathrm{complex}}$ corresponds to the potential energy of dipeptide-cation complex, 
$E_{\mathrm{dipeptide}}$ is the potential energy of the dipeptide alone fixed in the cation bound conformation, and $E_{\mathrm{cation}}$ is the potential energy of the isolated cation. 

Further data and properties can be extracted from the raw and normalized data \cite{Hu2021cation} that is available from the NOMAD Repository and Archive.\cite{draxl2019nomad} 
The data set was deposited as populated ontology in OWL format\cite{Hu2021ontology} in the EDMOND repository of the Max Planck Society.
The construction of the ontology is described in the following subsection.

\subsection*{Ontology construction}
Ontology construction is an iterative process involving many steps from defining common vocabularies, identifying the most important concepts and their relations to modelling such concepts in a semantically correct and still useful and applicable way.
They can be used to enrich, annotate and link data that is then called {\em linked data} and usually expressed in a triple format formed by \textit{subject-predicate-object}.\cite{al2020automatic}
In this work, the ontology builder Protégé\cite{musen2015protege} and the python package Owlready2\cite{lamy2017owlready} were employed to build ontologies in the OWL2 Web Ontology Language (\url{http://www.w3.org/TR/owl2-overview}) which is based on RDF -- the Resource Description Framework (\url{http://www.w3.org/TR/rdf-primer}).
Subjects and predicates are named using Internationalized Resource Identifiers (IRIs) (\url{https://tools.ietf.org/html/rfc3987}), while the object position can be filled by an IRI or a literal value (e.g. string or number). 
Ontologies created in this work have been tested with the OWL reasoner FACT++.\cite{tsarkov2006fact++}

\section*{Data Records}



Raw data and normalized data of the DFT calculations for this amino acid dipeptide data set is available from the NOMAD repository (\url{http://nomad-repository.eu}) via the DOI \href{https://dx.doi.org/10.17172/NOMAD/2021.02.10-1}{10.17172/NOMAD/2021.02.10-1}.\cite{Hu2021cation}
The NOMAD Archive contains all raw input, output, and property calculation files for download, while the NOMAD Repository contains normalized data, i.e. a digest of the DFT calculations. 
The extracted data in form of a populated ontology in OWL format is available download via the DOI \href{https://dx.doi.org/10.17617/3.5q}{10.17617/3.5q}.\cite{Hu2021ontology}
In the following two subsections, we briefly introduce the data and the concept of the provided ontology.

\subsection*{DFT data set}
\begin{figure}[ht] 
\vspace{0.0in}
\begin{center}
\includegraphics[height=0.8\textheight]{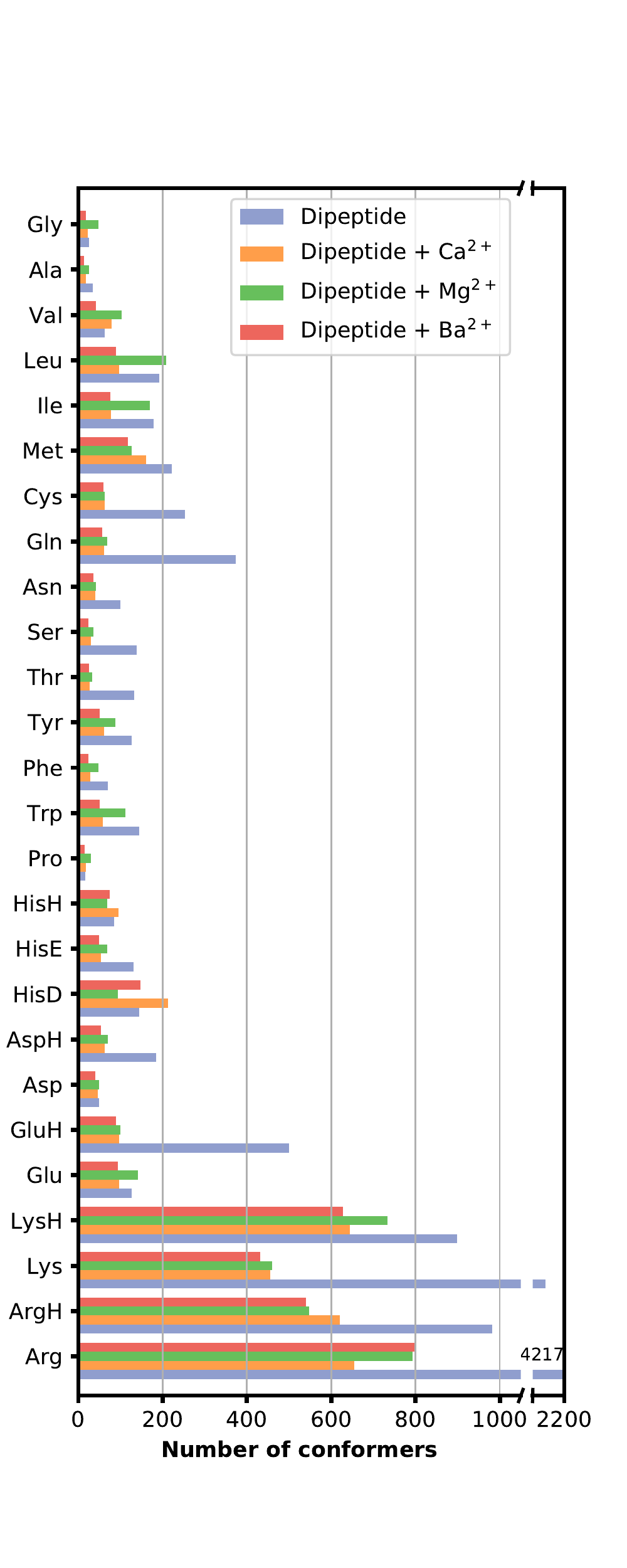}
\end{center}
\vspace{-0.2in}
\setlength{\belowcaptionskip}{0.2in}
\caption{Numbers of stationary points of each molecular system covered in this study.}
\label{num_conf}
\vspace{-0.2in}
\end{figure}

The distribution of the 21,909 stationary points of the amino acid dipeptide (plus cation) systems over the different amino acid building blocks is summarized in Figure~\ref{num_conf}.
This data is in particular intended for training energy functions in machine learning approaches in the context of force field development and parameterization.
Consequently, it consists not only of geometries with total energies for preferred low-energy conformers.
Instead, DFT-PBE+vdW calculations also included high-energy conformers.
The data we provide is particularly focused on parameterizing non-bonded interactions: 
    The above mentioned cation-peptide interaction energies were already used to tune force fields parameters of non-bonded interactions.\cite{wang2012development,amin2020benchmarking}
    The comparison to DFT-based vdW energies computed with the Tkatchenko-Scheffler formalism\cite{tkatchenko2009accurate} is useful to evaluate or adjust the non-bonded Lennard-Jones parameters $\epsilon$ and $\sigma$.
    Importantly, due to the spread over high and low energy conformations, diverse substructures and environments (due to cation binding), a range of partial charge values is sampled that informs about polarization and charge transfer.
    To that end, the electronic structure is simplified into partial charge models, based on Hirshfeld partitioning or Bader AIM analysis of the electron density. 
    The electron density, in combination with the nuclear charges, also defines the electrostatic potential (ESP) around the molecule, which can be used to derive force field parameters related to electrostatic interaction.\cite{li2017machine}
    The electron density has been used before to derive environment-specific force fields.\cite{cole2016biomolecular} Electron densities for a large set of molecules have been used to predict partial charges based on machine learning,\cite{rai2013fast, bleiziffer2018machine} to that end, an average over similar substructures in different molecules was used.

\begin{figure}[ht] 
\vspace{0.0in}
\begin{center}
\includegraphics[height=0.5\textheight]{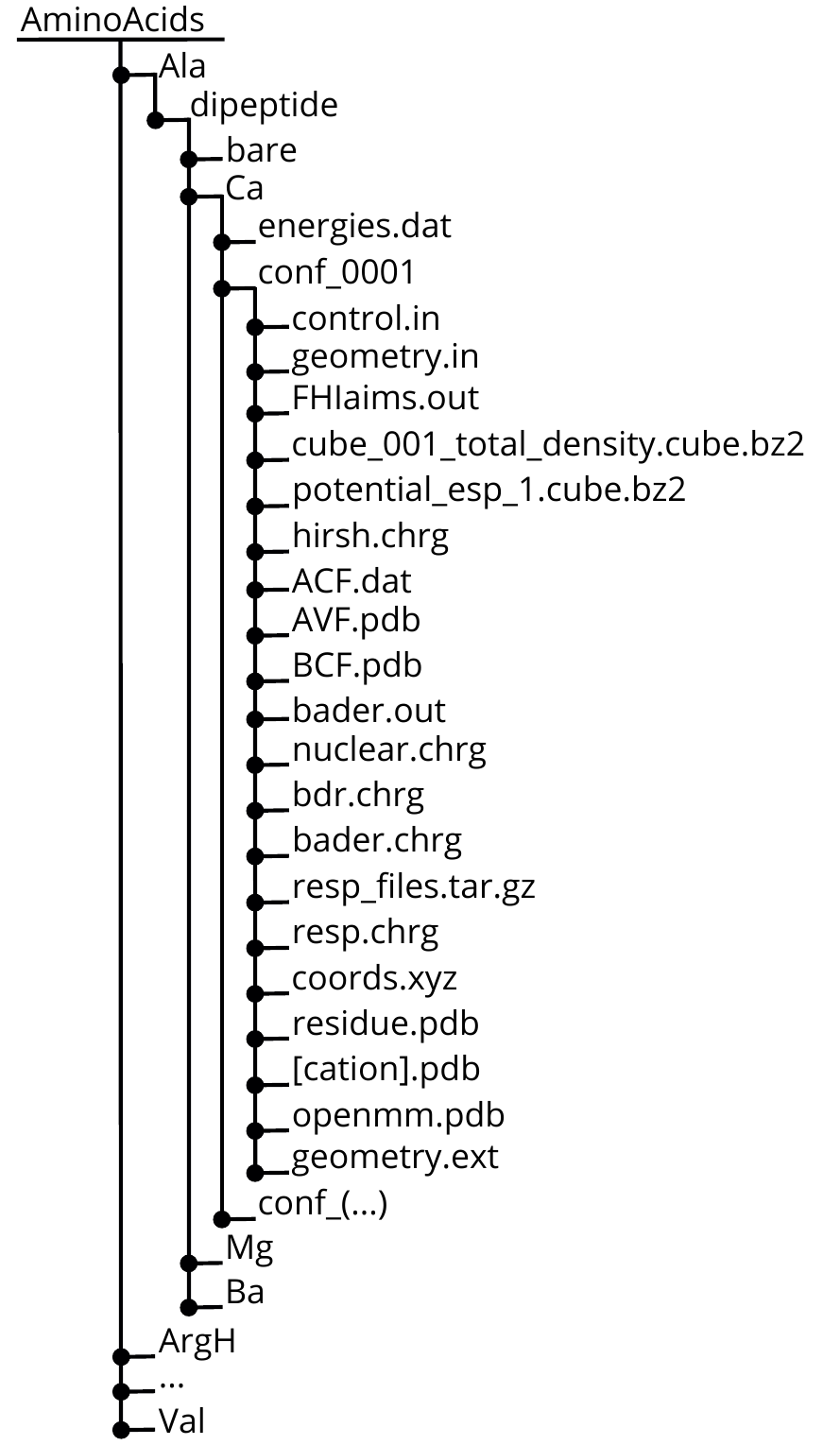}
\end{center}
\vspace{-0.2in}
\setlength{\belowcaptionskip}{0.2in}
\caption{Schematic representation of the folder structure of the data. Each folder, as exemplified for the Ca$^{2+}$-coordinated cysteine dipeptide, contains multiple properties per system.}
\label{tree}
\vspace{-0.2in}
\end{figure}

The data is first of all made available as a set of files. The different files, their content, and which programs to read or write them are given in Table~\ref{tab:filetypes}.
A direct way to access the data is to download the a compressed archive\cite{Hu2021cation} and browse the folder structure that is given in Figure~\ref{tree} or download from the same source the normalized data in json-files.
Attention, the download of the whole archive of raw data is about 1.5 TB in size (compressed).
This way of representing data however limits the automated access to the data by artificial agents or by researchers from other domain, as the metadata to the data is somewhat hidden. In order to alleviate this, the next section details the ontology which we developed in order to provide an extensible, machine-interpretable and machine-usable model for the automated access and post-processing of the data set.

\subsection*{Ontology}
\begin{figure}[ht] 
\vspace{0.0in}
\begin{center}
\includegraphics[width=0.35\textwidth]{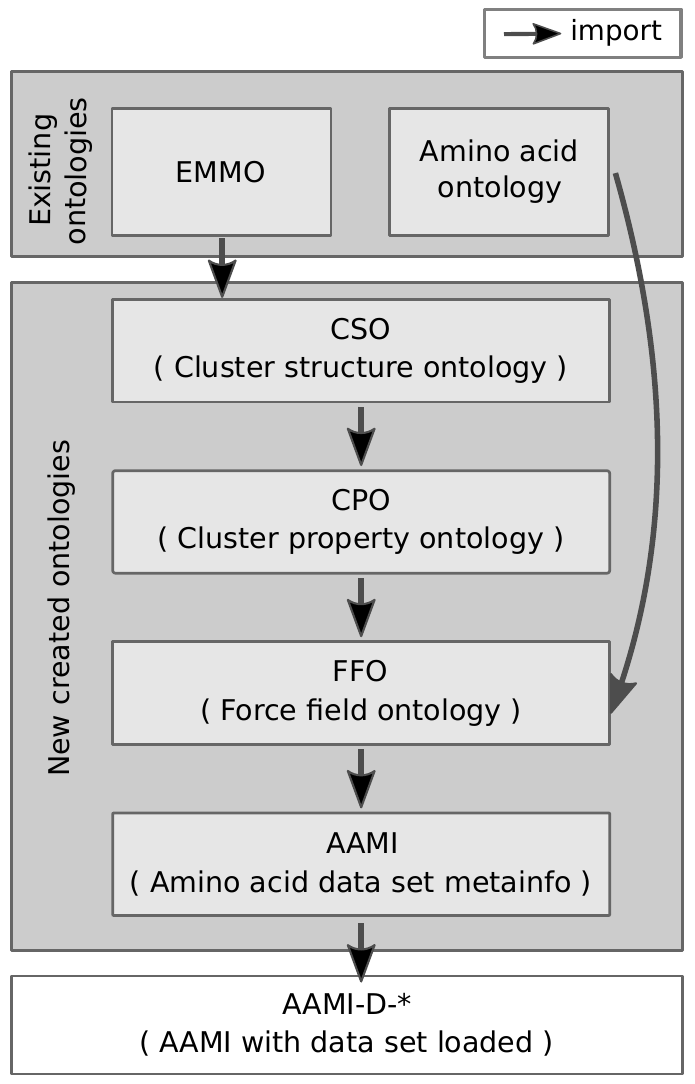}
\end{center}
\vspace{-0.2in}
\setlength{\belowcaptionskip}{0.2in}
\caption{Hierarchy of the ontologies linked to amino acid-cation meta-info (AAMI). Details of the ontologies and relations among them are described in Section~\textit{Ontology}}.
\label{Ontology}
\vspace{-0.2in}
\end{figure}

\begin{figure}[ht]
\vspace{0.0in}
\begin{center}
\includegraphics[width=0.7\textwidth]{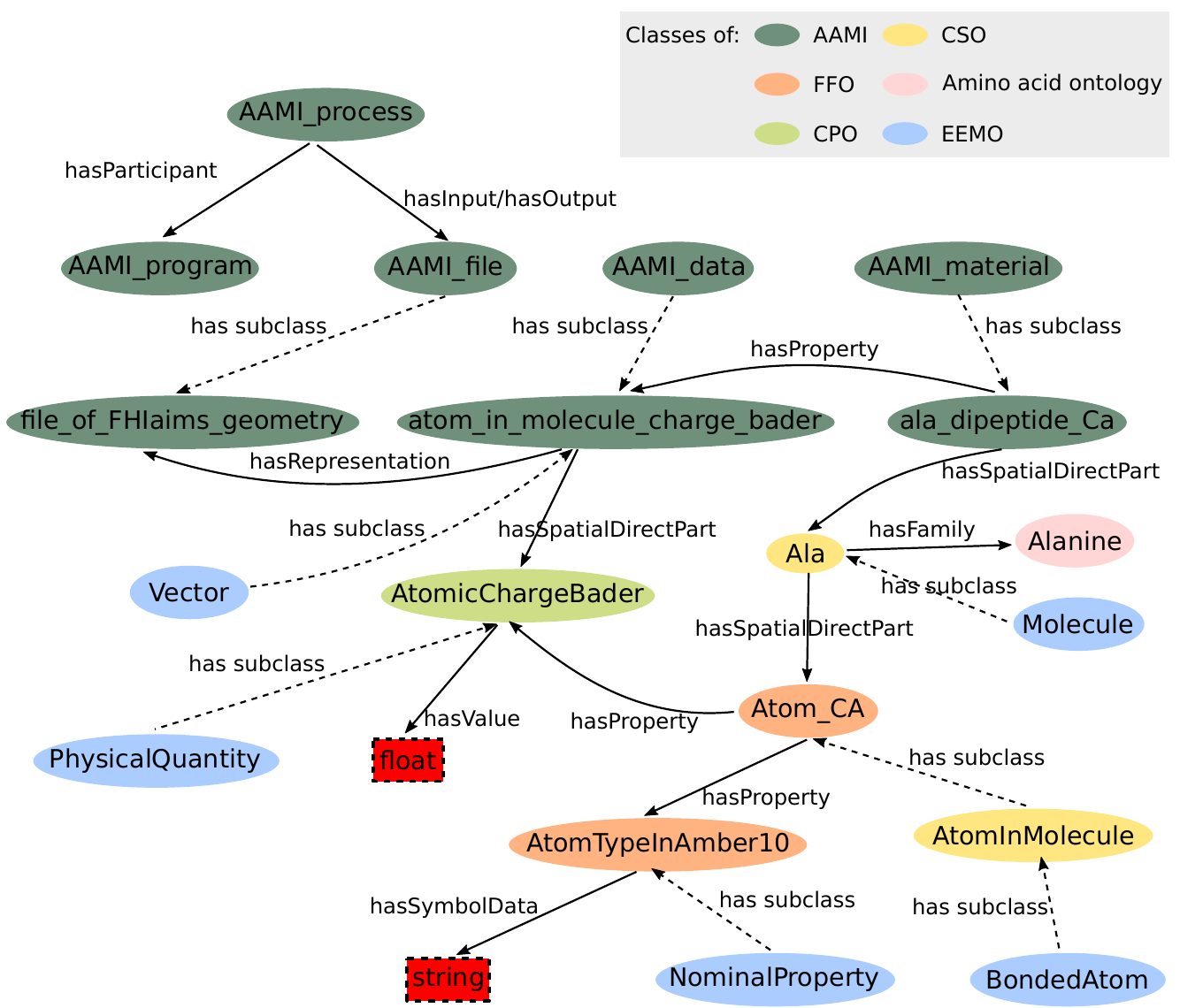}
\end{center}
\vspace{-0.2in}
\setlength{\belowcaptionskip}{0.2in}
\caption{Partial high-level class structure of AAMI ontology. Ovals represent classes, where classes from different ontologies are color coded. Rectangles represent literals. Solid lines are properties and dotted lines represent the relation of `\textit{has subclass}'. 
}
\label{Ontology_overview}
\vspace{-0.2in}
\end{figure}

AAMI (Amino Acid Meta-Info) is an ontology created ``bottom up'' to specifically represent the meta-information of this amino acid-cation data set in a machine-understandable and machine-processable way.
AAMI does not only contain metadata of properties, it also covers processes of analysis, such as participants in each process and their roles, which further makes data interpretable and understandable.
Two existing ontologies were re-used in AAMI: 
the European Materials Modelling Ontology (EMMO) (\url{https://emmc.info/emmo-info}), which provides a representational framework for materials modelling and characterization knowledge, and the Amino Acid Ontology (\url{http://bioportal.bioontology.org/ontologies/AMINO-ACID}), which provides structured knowledge of amino acids and their properties. 
By reusing existing terms in EMMO and Amino Acid Ontology rather than creating the ontology from scratch, terms in AAMI were connected to upper level concepts and can be potentially linked to further ontologies. 
Moreover, users are able to take advantage of data and annotations that are already used in those ontologies.
The primary aim of AAMI is to make our data set FAIR (Findable, Accessible, Interoperable, and Reusable),\cite{wilkinson2016fair} in particular accessible, interoperable and reusable. 
The elements of AAMI can be found in Figure~\ref{Ontology}; in the AAMI ecosystem, we created:

\begin{enumerate}
\item CSO (cluster structure ontology) represents concepts and relations for structure description of non-periodic system, EMMO was imported. 351 classes and 2053 axioms were created.

\item CPO (cluster property ontology) describes properties of non-periodic system. CSO was imported and 450 classes and 2984 axioms were created.

\item FFO (force field ontology) represents concepts in force fields, e.g. atom type and atom class. Amino acid ontology and CPO were imported. 563 classes and 4453 axioms were created.  

\item AAMI represents concepts and relations in the amino acids-cation data set. FFO was imported and 787 classes and 5466 axioms were created.

\item AAMI-D-* is the ontological representation of the data set in this study and was built from AAMI by populating with the data set. The data set was populated for a series of files, the asterisk represents the name of the corresponding amino acid, e.g. ALA, ARG \textit{etc.}

\end{enumerate}
Partial high level class organization and some of the classes and relations of AAMI are shown in Figure~\ref{Ontology_overview} to give an overview of the organization of the ontology and how terms from each ontology are related to each other. 
The primary use of AAMI is to annotate database records.
However, since ontologies in this work were developed with the OWL2 Web Ontology Language, which represents data by sets of subject-predicate-object statements, so called \textit{triples}, the underlying computational logic enables automatic inference and querying over data repositories.
In principle, any question framed in the respective mathematical logic can be answered in finite number of steps. 
However, such reasoning capabilities are currently limited to description logic.
Such data query can be done with the ontology and linked data query language, SPARQL (\url{https://www.w3.org/TR/sparql11-query}).
A user can query for sub-classes, relations between classes, functional annotation, and so on. 
We provide two sample queries in this work to guide users to build their own queries. 
Stardog Studio (\url{https://www.stardog.com/studio}) was used as \textit{triple store} and employed to perform the SPARQL queries. 
The main query form in SPARQL is a SELECT query. A SELECT query has two main components: a list of selected variables and a WHERE clause for specifying the graph patterns to match.
For example, according to the graph shown in Figure~\ref{Ontology_overview}, we can query for Bader charges of atoms which have atom type of ``1'' in Amber10 with a SELECT query as follows:

\begin{lstlisting}[language=SPARQL,breaklines]
PREFIX rdf: <http://www.w3.org/1999/02/22-rdf-syntax-ns#>
PREFIX owl: <http://www.w3.org/2002/07/owl#>
PREFIX rdfs: <http://www.w3.org/2000/01/rdf-schema#>
PREFIX xsd: <http://www.w3.org/2001/XMLSchema#>
PREFIX cso: <http://www.semanticweb.org/ClusterStructure.owl#>
PREFIX cpo: <http://www.semanticweb.org/ClusterProperty.owl#>
PREFIX ffo: <http://www.semanticweb.org/ForceField.owl#>
PREFIX aami: <http://www.semanticweb.org/AAMI.owl#>
PREFIX ala: <http://www.semanticweb.org/AAMI-D-Ala-Dipeptide.owl#>
PREFIX hasProperty:<http://emmo.info/emmo/middle/properties#EMMO_e1097637_70d2_4895_973f_2396f04fa204>
PREFIX hasSymbolData:<http://emmo.info/emmo/middle/perceptual#EMMO_23b579e1_8088_45b5_9975_064014026c42>

SELECT ?atom ?n
WHERE {
    ?atom hasProperty: ?atomtype.
    ?atomtype a ffo:AtomTypeInAmber10.
    ?atomtype hasSymbolData: "1"^^xsd:string.
    ?atom hasProperty: ?badercharge.
    ?badercharge a cpo:AtomicChargeBader.
    ?badercharge cpo:hasValue ?n
}
\end{lstlisting}
\noindent
AAMI-D-*, which is AAMI with data loaded, needs to be imported before the query.
The resulting list shows all atoms of type 1 in Amber10, \textit{i.e.} hydrogen atoms bound to a peptide bond nitrogen, and their Bader charges:
\begin{lstlisting}[language=SPARQL]
ala#Atom_HN_11_alaD_Ca_conf_0017  0.450512 
ala#Atom_HN_11_alaD_Ca_conf_0018  0.486539 
ala#Atom_HN_11_alaD_Ca_conf_0014  0.450169 
ala#Atom_HN_11_alaD_Ca_conf_0012  0.484383 
ala#Atom_HN_11_alaD_Ca_conf_0002  0.442222 
ala#Atom_HN_11_alaD_Ca_conf_0006  0.452150
...
\end{lstlisting}

Another useful query is DESCRIBE, which returns all the outgoing edges of a node. 
DESCRIBE is most useful when we don’t know much about the ontology and want to quickly see the terms used in the triples. 
For example, we can query ``describe individuals which belong to class Atom\_C'' with DESCRIBE query within AAMI-D-* data repositories with same PREFIX definition as in SELECT query:
\begin{lstlisting}[language=SPARQL]
DESCRIBE ?atom
WHERE {
    ?atom a ffo:Atom_C
}
\end{lstlisting}
In the following we display part of the output of the query, from which we can see that an individual  ``Atom\_C\_9\_alaD\_Ca\_conf\_0017'' belongs to class ``Atom\_C'' and has properties of ``AtomicChargeBader\_1.35427'', ``position9'' and so on.
\begin{lstlisting}[language=SPARQL]
@prefix ffo: <http://www.semanticweb.org/ForceField.owl#> .
@prefix ala: <http://www.semanticweb.org/AAMI-D-Ala-Dipeptide.owl#> .
@prefix hasProperty: <http://emmo.info/emmo/middle/properties#EMMO_e1097637_70d2_4895_973f_2396f04fa204> .

{
    <ala#Atom_C_9_alaD_Ca_conf_0017> a owl:NamedIndividual , ffo:Atom_C ;
      <hasProperty> <ala#AtomicChargeBader_1.35427> ,  <ala#EffectivePolarizability_9.839967844590108> ,  <ala#position9>, ...
}
\end{lstlisting}

With Stardog Studio, the result of such query can be written out in various file formats for further usage, e.g. \texttt{XML}, \texttt{JSON-LD} for triples output or \texttt{CSV} for tabular output.


\section*{Technical Validation}


The reliability of the DFT-PBE+vdW level of theory for amino acids and amino acids binding divalent cations was evaluated before.
In reference,\cite{ropo2016first} single-point energy calculations were performed on all structures of alanine (Ala) and phenylalanine (Phe) amino acids in isolation as well as binding with a Ca$^{2+}$ cation employing Møller-Plesset second-order perturbation theory (MP2),\cite{moller1934note, head1988mp2}
Both structural sets of Ala and Phe yielded a mean absolute error (MAE) within chemical accuracy of 1 kcal/mol with the theory of PBE+vdW.
A different long-range dispersion method, many-body dispersion model (PBE+MBD),\cite{ambrosetti2014long} didn't show significant improvement for isolated amino acids. Also with a more expensive hybrid exchange correlation functional, PBE0 (PBE0+MBD),\cite{ambrosetti2014long} MAEs were not significantly improved.
However, maximum error of Phe was reduced from 2 kcal/mol to 1.3 kcal/mol. 
MAEs were slightly higher with PBE+vdW when Ca$^{2+}$ was involved.
They reached 1 kcal/mol and 2 kcal/mol for Ala + Ca$^{2+}$ and Phe + Ca$^{2+}$, respectively. 
Employing both, many-body dispersion and the hybrid functional PBE0, only improved the MAE to about 1 kcal/mol.
PBE+vdW represents a good compromise between accuracy and computational cost.
In a manuscript on histidine-zinc interactions,\cite{schneider2018relative} DLPNO-CCSD(T)\cite{riplinger2013efficient, riplinger2013natural} was employed to benchmark several DFAs as well as the wave function-based MP2 method. 
The evaluated systems are 
(a)~negatively charged acetylhistidine (AcH) with and without a Zn$^{2+}$ cation, and 
(b)~neutral AcH with and without a Zn$^{2+}$ cation. 
The results showed that PBE+vdW gave an acceptable accuracy. 
As a conclusion, PBE+vdW is an excellent starting point for more exhaustive future benchmark work of new electronic structure methods for cation-peptide systems.

The validation of the sampling method can be elucidated by the work in reference.\cite{supady2015first} 
Genetic algorithm was employed to do the sampling of the low-energy segment on the conformation space of seven dipeptides: Glycine (Gly), Alanine (Ala), Phenylalanine (Phe), Valine (Val), Tryptophan (Trp), Leucine (Leu), Isoleucine (Ile) in reference.\cite{supady2015first}
Conformers in our previous data set\cite{ropo2016first} were used as references.

The potential usage of our data set has been confirmed in reference.\cite{amin2020benchmarking}
In this work, our data set was used to assess the accuracy of existing FFs by their abilities to reproduce quantum mechanical (QM) interaction energies of Ca$^{2+}$-dipeptide. By relating the parameter space to conformational space, the utility of our data set as a reference for future optimization of polarizable force fields is illustrated.

\begin{figure}[ht]
\vspace{0.0in}
\begin{center}
\subfigure[\textbf{Dipeptide bare}]{\includegraphics[height=0.35\textheight]{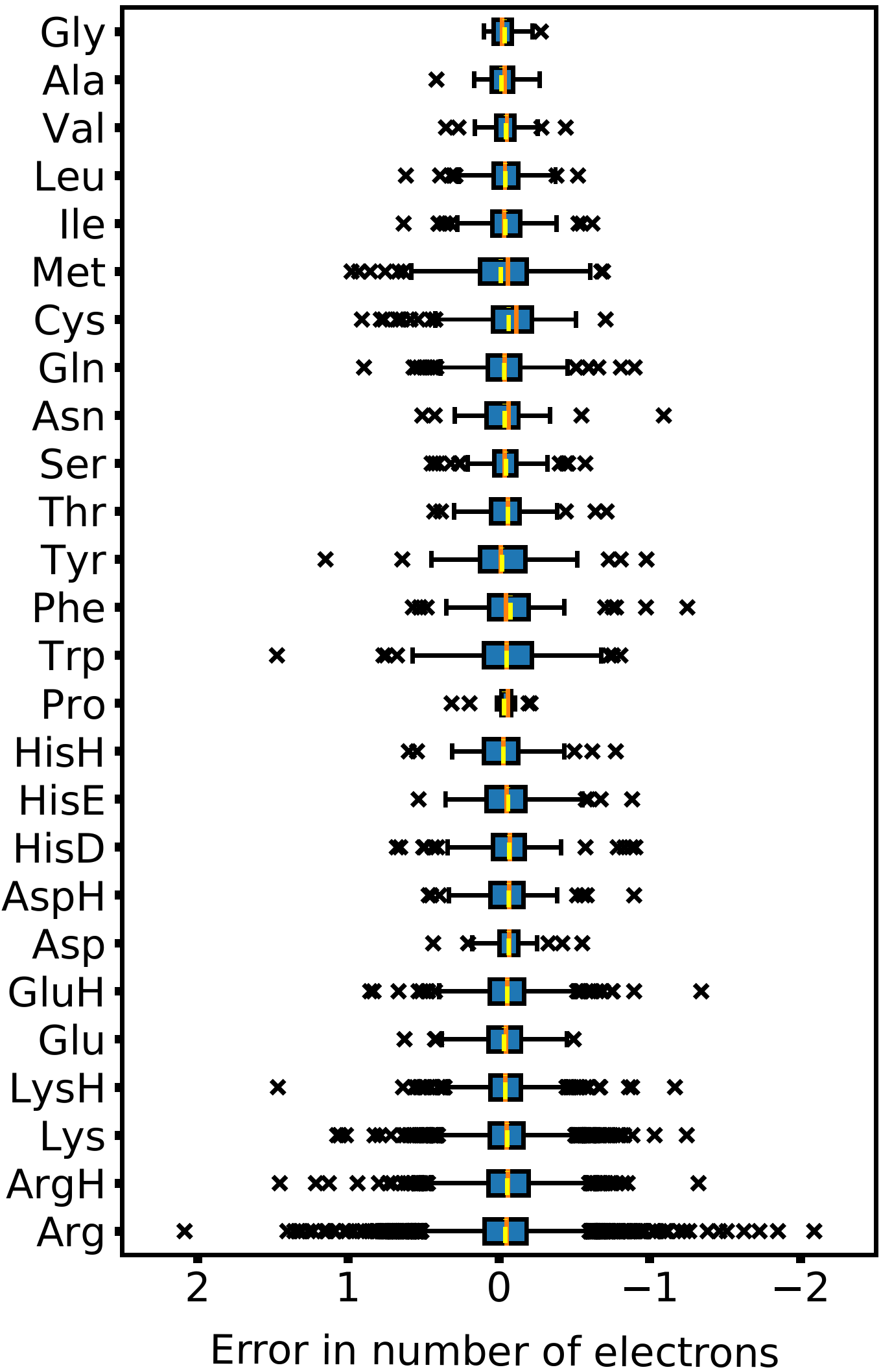}}
\subfigure[\textbf{Dipeptide+Ca$^{2+}$}]{\includegraphics[height=0.35\textheight]{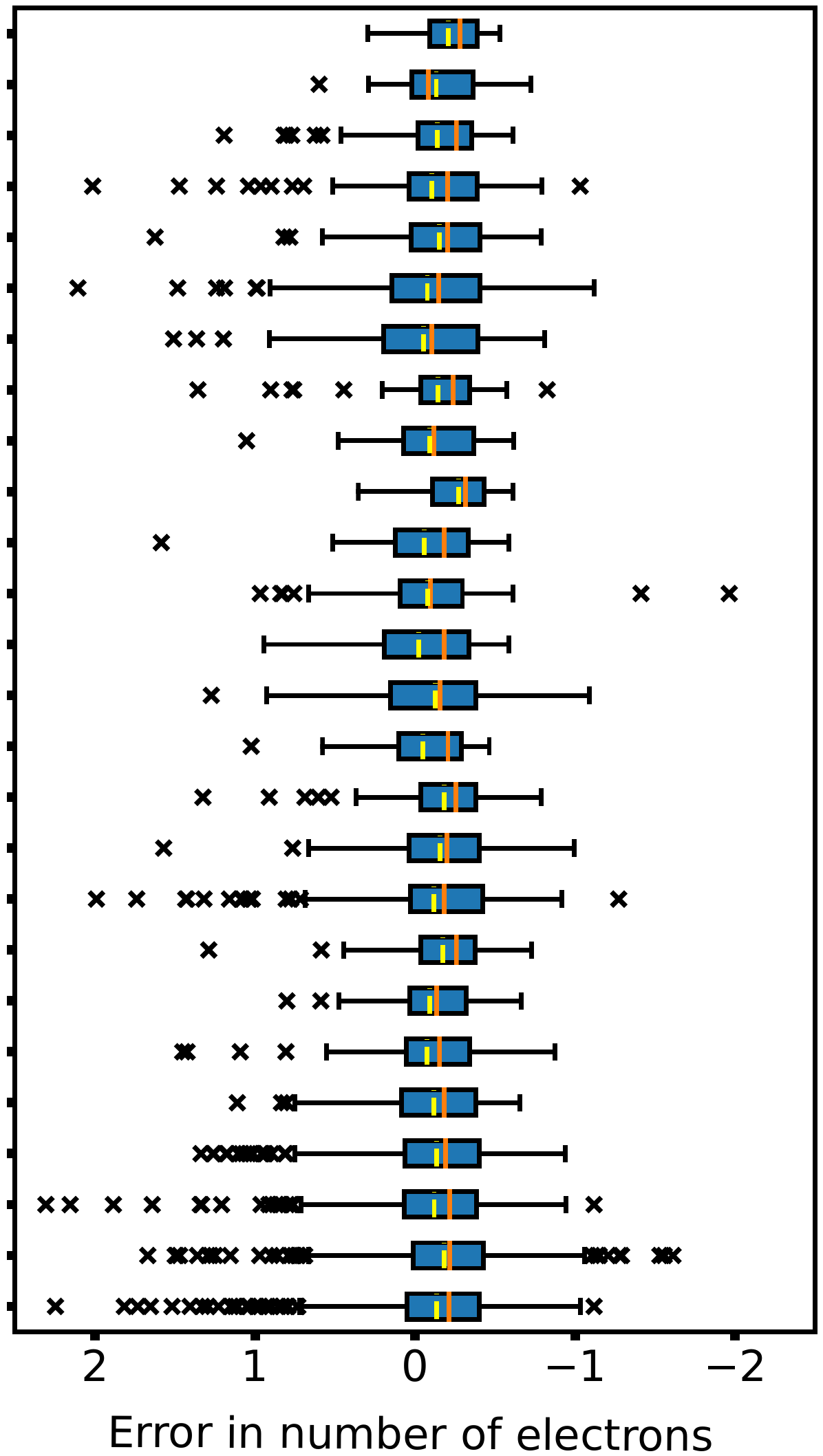}}
\subfigure[\textbf{Dipeptide+Mg$^{2+}$}]{\includegraphics[height=0.35\textheight]{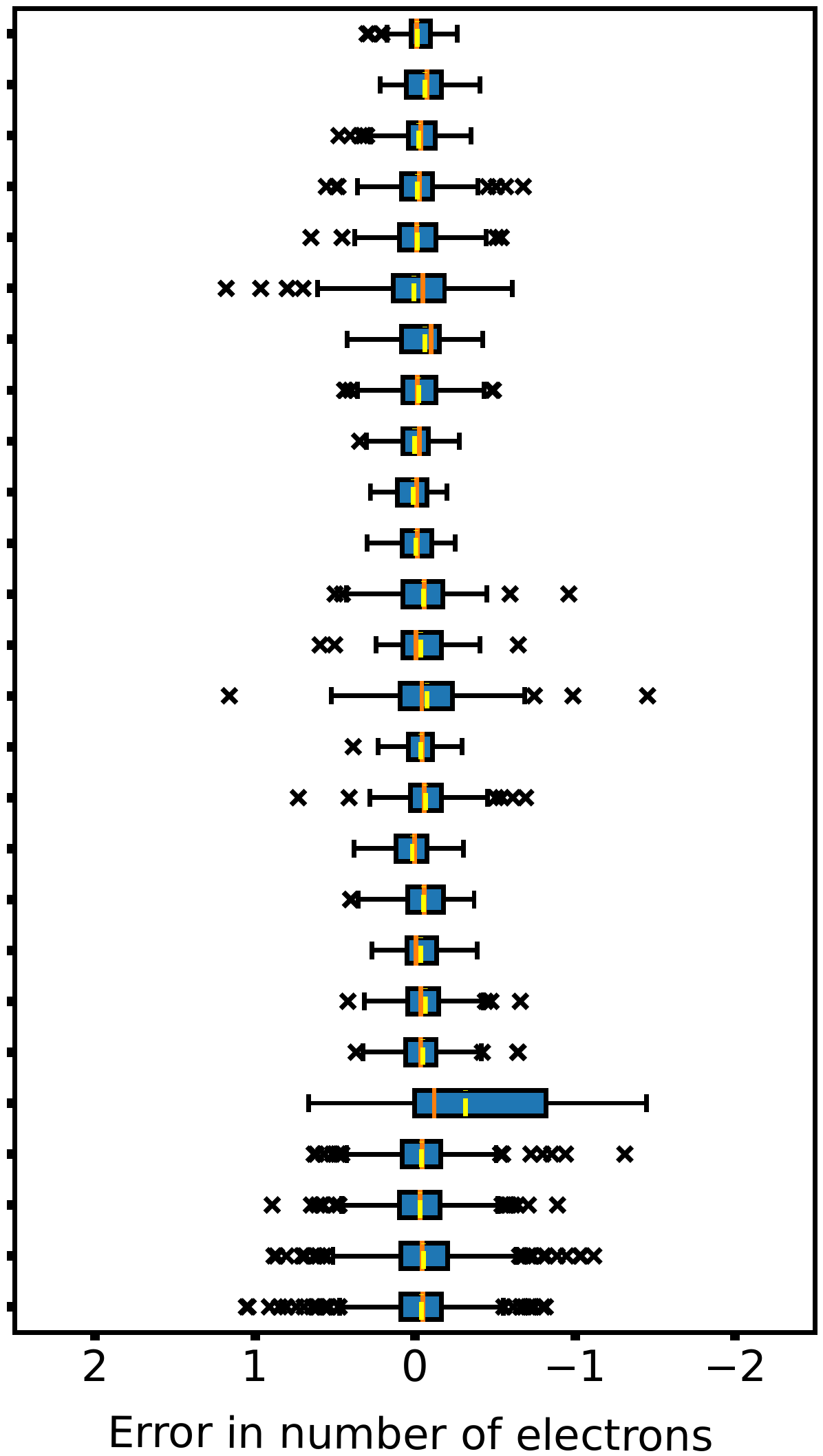}}
\end{center}
\vspace{-0.1in}
\setlength{\belowcaptionskip}{0.2in}
\caption{Error in numbers of electrons from Bader analysis of Dipeptide (a)~bare, (b)~with Ca$^{2+}$ and (c)~with Mg$^{2+}$. The upper and lower lines of the rectangles mark the 75\% and 25\% percentiles of the distribution, the orange and yellow horizontal lines in the box indicate the median (50\% percentile) and mean value, and the upper and lower lines of the “error bars” depict the 99\% and 1\% percentiles. Crosses represent the outliers. 
}
\label{num_electron}
\vspace{-0.2in}
\end{figure}

An assessment of the reliability of Bader charge analysis of bare dipeptides as well as dipeptide-Ca$^{2+}$ and dipeptide-Mg$^{2+}$ complexes is shown in Figure~\ref{num_electron}. 
The number of electrons from Bader charge analysis yielded high errors in some structures of dipeptide-Ca$^{2+}$, reaching 2 electrons.
This error apparently results from too wide grid spacing at regions of rapid density change (near ``heavy'' cores) when writing the electron density to cube files, the input for the Bader analysis code. 
Changes of electron density are particularly large close to the cations in the investigated clusters, so in principle grid spacings adjusted to the respective systems would be required.
Overall however, the mean errors of each amino acids are around 0. 
The errors of dipeptide-Mg$^{2+}$ have the same trend, but are smaller than the errors of dipeptide-Ca$^{2+}$ due to the smaller radius of Mg$^{2+}$. 
Ba$^{2+}$ is much heavier than Ca$^{2+}$ and Mg$^{2+}$, the rise in density close to the atomic center is much steeper. 
To analyse the Bader charges of dipeptide-Ba$^{2+}$ complexes, a much smaller grid spacing is needed. 
However, this will result in impractically large cube files of electron density.
So in this work, we did not present the electron density and Bader charges of dipeptide-Ba$^{2+}$.

\section*{Usage Notes}



Structures in this data set are stationary-point geometries, most of them can be expected to be minima, yet there are certainly also saddle points. 
All files in NOMAD repository can be downloaded through \texttt{curl} based on upload and entry IDs (variables: \texttt{upload\_id} and \texttt{entry\_id} below). The command below downloads all files in one calculation: 
\begin{lstlisting}[language=SPARQL,breaklines]
curl "http://repository.nomad-coe.eu/app/api/raw/calc/upload_id/entry_id/*" -o download.zip
\end{lstlisting}
Some metadata about the DFT calculations can be browsed at the NOMAD Archive page (\url{https://www.nomad-coe.eu/the-project/nomad-archive/archive-meta-info}).
There are numerous tools to perform SPARQL queries, e.g. Protégé,\cite{musen2015protege} RDFLib (\url{https://github.com/RDFLib/rdflib}), Apache Jena (\url{https://jena.apache.org}), and so on.

\section*{Code availability} \label{custom_codes}

All custom codes used in this study have been uploaded to Github and can be found here:\\ \url{https://github.com/XiaojuanHu/AA\_property\_calculation}.

\newpage


\begin{thebibliography}{10}
\urlstyle{rm}
\expandafter\ifx\csname url\endcsname\relax
  \def\url#1{\texttt{#1}}\fi
\expandafter\ifx\csname urlprefix\endcsname\relax\def\urlprefix{URL }\fi
\expandafter\ifx\csname doiprefix\endcsname\relax\def\doiprefix{DOI: }\fi
\providecommand{\bibinfo}[2]{#2}
\providecommand{\eprint}[2][]{\url{#2}}

\bibitem{permyakov2009metalloproteomics}
\bibinfo{author}{Permyakov, E.}
\newblock \emph{\bibinfo{title}{Metalloproteomics}}, vol.~\bibinfo{volume}{2}
  (\bibinfo{publisher}{John Wiley \& Sons}, \bibinfo{year}{2009}).

\bibitem{bertini2007biological}
\bibinfo{author}{Bertini, G.} \emph{et~al.}
\newblock \emph{\bibinfo{title}{Biological inorganic chemistry: structure and
  reactivity}} (\bibinfo{publisher}{University Science Books},
  \bibinfo{year}{2007}).

\bibitem{tamames2007analysis}
\bibinfo{author}{Tamames, B.}, \bibinfo{author}{Sousa, S.~F.},
  \bibinfo{author}{Tamames, J.}, \bibinfo{author}{Fernandes, P.~A.} \&
  \bibinfo{author}{Ramos, M.~J.}
\newblock \bibinfo{journal}{\bibinfo{title}{Analysis of zinc-ligand bond
  lengths in metalloproteins: trends and patterns}}.
\newblock {\emph{\JournalTitle{Proteins: Structure, Function, and
  Bioinformatics}}} \textbf{\bibinfo{volume}{69}}, \bibinfo{pages}{466--475}
  (\bibinfo{year}{2007}).

\bibitem{sala2018molecular}
\bibinfo{author}{Sala, D.}, \bibinfo{author}{Giachetti, A.} \&
  \bibinfo{author}{Rosato, A.}
\newblock \bibinfo{journal}{\bibinfo{title}{Molecular dynamics simulations of
  metalloproteins: A folding study of rubredoxin from {P}yrococcus furiosus}}.
\newblock {\emph{\JournalTitle{AIMS Biophys}}} \textbf{\bibinfo{volume}{5}},
  \bibinfo{pages}{77--96} (\bibinfo{year}{2018}).

\bibitem{zhou2011novel}
\bibinfo{author}{Zhou, M.} \emph{et~al.}
\newblock \bibinfo{journal}{\bibinfo{title}{A novel calcium-binding site of von
  {W}illebrand factor {A}2 domain regulates its cleavage by {ADAMTS}13}}.
\newblock {\emph{\JournalTitle{Blood}}} \textbf{\bibinfo{volume}{117}},
  \bibinfo{pages}{4623--4631} (\bibinfo{year}{2011}).

\bibitem{gogoi2016heterogeneous}
\bibinfo{author}{Gogoi, P.}, \bibinfo{author}{Chandravanshi, M.},
  \bibinfo{author}{Mandal, S.~K.}, \bibinfo{author}{Srivastava, A.} \&
  \bibinfo{author}{Kanaujia, S.~P.}
\newblock \bibinfo{journal}{\bibinfo{title}{Heterogeneous behavior of
  metalloproteins toward metal ion binding and selectivity: insights from
  molecular dynamics studies}}.
\newblock {\emph{\JournalTitle{Journal of Biomolecular Structure and
  Dynamics}}} \textbf{\bibinfo{volume}{34}}, \bibinfo{pages}{1470--1485}
  (\bibinfo{year}{2016}).

\bibitem{baldauf2013cations}
\bibinfo{author}{Baldauf, C.} \emph{et~al.}
\newblock \bibinfo{journal}{\bibinfo{title}{How cations change peptide
  structure}}.
\newblock {\emph{\JournalTitle{Chemistry--A European Journal}}}
  \textbf{\bibinfo{volume}{19}}, \bibinfo{pages}{11224--11234}
  (\bibinfo{year}{2013}).

\bibitem{de2017mapping}
\bibinfo{author}{De, S.}, \bibinfo{author}{Musil, F.}, \bibinfo{author}{Ingram,
  T.}, \bibinfo{author}{Baldauf, C.} \& \bibinfo{author}{Ceriotti, M.}
\newblock \bibinfo{journal}{\bibinfo{title}{Mapping and classifying molecules
  from a high-throughput structural database}}.
\newblock {\emph{\JournalTitle{Journal of Cheminformatics}}}
  \textbf{\bibinfo{volume}{9}}, \bibinfo{pages}{1--14} (\bibinfo{year}{2017}).

\bibitem{ropo2016trends}
\bibinfo{author}{Ropo, M.}, \bibinfo{author}{Blum, V.} \&
  \bibinfo{author}{Baldauf, C.}
\newblock \bibinfo{journal}{\bibinfo{title}{Trends for isolated amino acids and
  dipeptides: {C}onformation, divalent ion binding, and remarkable similarity
  of binding to calcium and lead}}.
\newblock {\emph{\JournalTitle{Scientific Reports}}}
  \textbf{\bibinfo{volume}{6}}, \bibinfo{pages}{1--11} (\bibinfo{year}{2016}).

\bibitem{vitalini2015dynamic}
\bibinfo{author}{Vitalini, F.}, \bibinfo{author}{Mey, A.~S.},
  \bibinfo{author}{No{\'e}, F.} \& \bibinfo{author}{Keller, B.~G.}
\newblock \bibinfo{journal}{\bibinfo{title}{Dynamic properties of force
  fields}}.
\newblock {\emph{\JournalTitle{The Journal of Chemical Physics}}}
  \textbf{\bibinfo{volume}{142}}, \bibinfo{pages}{02B611\_1}
  (\bibinfo{year}{2015}).

\bibitem{schneider2018relative}
\bibinfo{author}{Schneider, M.} \& \bibinfo{author}{Baldauf, C.}
\newblock \bibinfo{journal}{\bibinfo{title}{Relative energetics of
  acetyl-histidine protomers with and without {Z}n$^{2+}$ and a benchmark of
  energy methods}}.
\newblock {\emph{\JournalTitle{arXiv preprint arXiv:1810.10596}}}
  (\bibinfo{year}{2018}).

\bibitem{maksimov2021conformational}
\bibinfo{author}{Maksimov, D.}, \bibinfo{author}{Baldauf, C.} \&
  \bibinfo{author}{Rossi, M.}
\newblock \bibinfo{journal}{\bibinfo{title}{The conformational space of a
  flexible amino acid at metallic surfaces}}.
\newblock {\emph{\JournalTitle{International Journal of Quantum Chemistry}}}
  \textbf{\bibinfo{volume}{121}}, \bibinfo{pages}{e26369}
  (\bibinfo{year}{2021}).

\bibitem{marianski2016assessing}
\bibinfo{author}{Marianski, M.}, \bibinfo{author}{Supady, A.},
  \bibinfo{author}{Ingram, T.}, \bibinfo{author}{Schneider, M.} \&
  \bibinfo{author}{Baldauf, C.}
\newblock \bibinfo{journal}{\bibinfo{title}{Assessing the accuracy of
  across-the-scale methods for predicting carbohydrate conformational energies
  for the examples of glucose and $\alpha$-maltose}}.
\newblock {\emph{\JournalTitle{Journal of Chemical Theory and Computation}}}
  \textbf{\bibinfo{volume}{12}}, \bibinfo{pages}{6157--6168}
  (\bibinfo{year}{2016}).

\bibitem{wang2001automatic}
\bibinfo{author}{Wang, J.} \& \bibinfo{author}{Kollman, P.~A.}
\newblock \bibinfo{journal}{\bibinfo{title}{Automatic parameterization of force
  field by systematic search and genetic algorithms}}.
\newblock {\emph{\JournalTitle{Journal of Computational Chemistry}}}
  \textbf{\bibinfo{volume}{22}}, \bibinfo{pages}{1219--1228}
  (\bibinfo{year}{2001}).

\bibitem{oostenbrink2004biomolecular}
\bibinfo{author}{Oostenbrink, C.}, \bibinfo{author}{Villa, A.},
  \bibinfo{author}{Mark, A.~E.} \& \bibinfo{author}{Van~Gunsteren, W.~F.}
\newblock \bibinfo{journal}{\bibinfo{title}{A biomolecular force field based on
  the free enthalpy of hydration and solvation: the {GROMOS} force-field
  parameter sets 53{A}5 and 53{A}6}}.
\newblock {\emph{\JournalTitle{Journal of Computational Chemistry}}}
  \textbf{\bibinfo{volume}{25}}, \bibinfo{pages}{1656--1676}
  (\bibinfo{year}{2004}).

\bibitem{wang2000well}
\bibinfo{author}{Wang, J.}, \bibinfo{author}{Cieplak, P.} \&
  \bibinfo{author}{Kollman, P.~A.}
\newblock \bibinfo{journal}{\bibinfo{title}{How well does a restrained
  electrostatic potential ({RESP}) model perform in calculating conformational
  energies of organic and biological molecules?}}
\newblock {\emph{\JournalTitle{Journal of Computational Chemistry}}}
  \textbf{\bibinfo{volume}{21}}, \bibinfo{pages}{1049--1074}
  (\bibinfo{year}{2000}).

\bibitem{riniker2018fixed}
\bibinfo{author}{Riniker, S.}
\newblock \bibinfo{journal}{\bibinfo{title}{Fixed-charge atomistic force fields
  for molecular dynamics simulations in the condensed phase: {A}n overview}}.
\newblock {\emph{\JournalTitle{Journal of Chemical Information and Modeling}}}
  \textbf{\bibinfo{volume}{58}}, \bibinfo{pages}{565--578}
  (\bibinfo{year}{2018}).

\bibitem{allen2004energetics}
\bibinfo{author}{Allen, T.~W.}, \bibinfo{author}{Andersen, O.~S.} \&
  \bibinfo{author}{Roux, B.}
\newblock \bibinfo{journal}{\bibinfo{title}{Energetics of ion conduction
  through the gramicidin channel}}.
\newblock {\emph{\JournalTitle{Proceedings of the National Academy of
  Sciences}}} \textbf{\bibinfo{volume}{101}}, \bibinfo{pages}{117--122}
  (\bibinfo{year}{2004}).

\bibitem{roca2003theoretical}
\bibinfo{author}{Roca, M.} \emph{et~al.}
\newblock \bibinfo{journal}{\bibinfo{title}{Theoretical modeling of enzyme
  catalytic power: analysis of “cratic” and electrostatic factors in
  catechol {O}-methyltransferase}}.
\newblock {\emph{\JournalTitle{Journal of the American Chemical Society}}}
  \textbf{\bibinfo{volume}{125}}, \bibinfo{pages}{7726--7737}
  (\bibinfo{year}{2003}).

\bibitem{zeng2013f130l}
\bibinfo{author}{Zeng, J.}, \bibinfo{author}{Jia, X.}, \bibinfo{author}{Zhang,
  J.~Z.} \& \bibinfo{author}{Mei, Y.}
\newblock \bibinfo{journal}{\bibinfo{title}{The {F}130{L} mutation in
  streptavidin reduces its binding affinity to biotin through electronic
  polarization effect}}.
\newblock {\emph{\JournalTitle{Journal of Computational Chemistry}}}
  \textbf{\bibinfo{volume}{34}}, \bibinfo{pages}{2677--2686}
  (\bibinfo{year}{2013}).

\bibitem{li2011structure}
\bibinfo{author}{Li, Y.~L.}, \bibinfo{author}{Mei, Y.}, \bibinfo{author}{Zhang,
  D.~W.}, \bibinfo{author}{Xie, D.~Q.} \& \bibinfo{author}{Zhang, J.~Z.}
\newblock \bibinfo{journal}{\bibinfo{title}{Structure and dynamics of a dizinc
  metalloprotein: effect of charge transfer and polarization}}.
\newblock {\emph{\JournalTitle{The Journal of Physical Chemistry B}}}
  \textbf{\bibinfo{volume}{115}}, \bibinfo{pages}{10154--10162}
  (\bibinfo{year}{2011}).

\bibitem{xie2009coupled}
\bibinfo{author}{Xie, W.}, \bibinfo{author}{Pu, J.} \& \bibinfo{author}{Gao,
  J.}
\newblock \bibinfo{journal}{\bibinfo{title}{A coupled polarization-matrix
  inversion and iteration approach for accelerating the dipole convergence in a
  polarizable potential function}}.
\newblock {\emph{\JournalTitle{The Journal of Physical Chemistry A}}}
  \textbf{\bibinfo{volume}{113}}, \bibinfo{pages}{2109--2116}
  (\bibinfo{year}{2009}).

\bibitem{ngo2015quantum}
\bibinfo{author}{Ngo, V.} \emph{et~al.}
\newblock \bibinfo{journal}{\bibinfo{title}{Quantum effects in cation
  interactions with first and second coordination shell ligands in
  metalloproteins}}.
\newblock {\emph{\JournalTitle{Journal of Chemical Theory and Computation}}}
  \textbf{\bibinfo{volume}{11}}, \bibinfo{pages}{4992--5001}
  (\bibinfo{year}{2015}).

\bibitem{amin2020benchmarking}
\bibinfo{author}{Amin, K.~S.} \emph{et~al.}
\newblock \bibinfo{journal}{\bibinfo{title}{Benchmarking polarizable and
  non-polarizable force fields for {C}a$^{2+}$--peptides against a
  comprehensive {QM} dataset}}.
\newblock {\emph{\JournalTitle{The Journal of Chemical Physics}}}
  \textbf{\bibinfo{volume}{153}}, \bibinfo{pages}{144102}
  (\bibinfo{year}{2020}).

\bibitem{liang1996parameter}
\bibinfo{author}{Liang, G.}, \bibinfo{author}{Fox, P.~C.} \&
  \bibinfo{author}{Bowen, J.~P.}
\newblock \bibinfo{journal}{\bibinfo{title}{Parameter analysis and refinement
  toolkit system and its application in {MM}3 parameterization for phosphine
  and its derivatives}}.
\newblock {\emph{\JournalTitle{Journal of Computational Chemistry}}}
  \textbf{\bibinfo{volume}{17}}, \bibinfo{pages}{940--953}
  (\bibinfo{year}{1996}).

\bibitem{faller1999automatic}
\bibinfo{author}{Faller, R.}, \bibinfo{author}{Schmitz, H.},
  \bibinfo{author}{Biermann, O.} \& \bibinfo{author}{M{\"u}ller-Plathe, F.}
\newblock \bibinfo{journal}{\bibinfo{title}{Automatic parameterization of force
  fields for liquids by simplex optimization}}.
\newblock {\emph{\JournalTitle{Journal of Computational Chemistry}}}
  \textbf{\bibinfo{volume}{20}}, \bibinfo{pages}{1009--1017}
  (\bibinfo{year}{1999}).

\bibitem{cisneros2014classical}
\bibinfo{author}{Cisneros, G.~A.}, \bibinfo{author}{Karttunen, M.},
  \bibinfo{author}{Ren, P.} \& \bibinfo{author}{Sagui, C.}
\newblock \bibinfo{journal}{\bibinfo{title}{Classical electrostatics for
  biomolecular simulations}}.
\newblock {\emph{\JournalTitle{Chemical Reviews}}}
  \textbf{\bibinfo{volume}{114}}, \bibinfo{pages}{779--814}
  (\bibinfo{year}{2014}).

\bibitem{smith2017ani}
\bibinfo{author}{Smith, J.~S.}, \bibinfo{author}{Isayev, O.} \&
  \bibinfo{author}{Roitberg, A.~E.}
\newblock \bibinfo{journal}{\bibinfo{title}{{ANI}-1: an extensible neural
  network potential with {DFT} accuracy at force field computational cost}}.
\newblock {\emph{\JournalTitle{Chemical Science}}}
  \textbf{\bibinfo{volume}{8}}, \bibinfo{pages}{3192--3203}
  (\bibinfo{year}{2017}).

\bibitem{rezac2018toward}
\bibinfo{author}{Rezac, J.}, \bibinfo{author}{B{\'\i}m, D.},
  \bibinfo{author}{Gutten, O.} \& \bibinfo{author}{Rulisek, L.}
\newblock \bibinfo{journal}{\bibinfo{title}{Toward accurate conformational
  energies of smaller peptides and medium-sized macrocycles: {MPCONF}196
  benchmark energy data set}}.
\newblock {\emph{\JournalTitle{Journal of Chemical Theory and Computation}}}
  \textbf{\bibinfo{volume}{14}}, \bibinfo{pages}{1254--1266}
  (\bibinfo{year}{2018}).

\bibitem{jurevcka2006benchmark}
\bibinfo{author}{Jure{\v{c}}ka, P.}, \bibinfo{author}{{\v{S}}poner, J.},
  \bibinfo{author}{{\v{C}}ern{\`y}, J.} \& \bibinfo{author}{Hobza, P.}
\newblock \bibinfo{journal}{\bibinfo{title}{Benchmark database of accurate
  ({MP}2 and {CCSD} ({T}) complete basis set limit) interaction energies of
  small model complexes, {DNA} base pairs, and amino acid pairs}}.
\newblock {\emph{\JournalTitle{Physical Chemistry Chemical Physics}}}
  \textbf{\bibinfo{volume}{8}}, \bibinfo{pages}{1985--1993}
  (\bibinfo{year}{2006}).

\bibitem{goerigk2017look}
\bibinfo{author}{Goerigk, L.} \emph{et~al.}
\newblock \bibinfo{journal}{\bibinfo{title}{A look at the density functional
  theory zoo with the advanced {GMTKN}55 database for general main group
  thermochemistry, kinetics and noncovalent interactions}}.
\newblock {\emph{\JournalTitle{Physical Chemistry Chemical Physics}}}
  \textbf{\bibinfo{volume}{19}}, \bibinfo{pages}{32184--32215}
  (\bibinfo{year}{2017}).

\bibitem{dohm2018comprehensive}
\bibinfo{author}{Dohm, S.}, \bibinfo{author}{Hansen, A.},
  \bibinfo{author}{Steinmetz, M.}, \bibinfo{author}{Grimme, S.} \&
  \bibinfo{author}{Checinski, M.~P.}
\newblock \bibinfo{journal}{\bibinfo{title}{Comprehensive thermochemical
  benchmark set of realistic closed-shell metal organic reactions}}.
\newblock {\emph{\JournalTitle{Journal of Chemical Theory and Computation}}}
  \textbf{\bibinfo{volume}{14}}, \bibinfo{pages}{2596--2608}
  (\bibinfo{year}{2018}).

\bibitem{yu2009extensive}
\bibinfo{author}{Yu, W.} \emph{et~al.}
\newblock \bibinfo{journal}{\bibinfo{title}{Extensive conformational searches
  of 13 representative dipeptides and an efficient method for dipeptide
  structure determinations based on amino acid conformers}}.
\newblock {\emph{\JournalTitle{Journal of Computational Chemistry}}}
  \textbf{\bibinfo{volume}{30}}, \bibinfo{pages}{2105--2121}
  (\bibinfo{year}{2009}).

\bibitem{kishor2008structural}
\bibinfo{author}{Kishor, S.}, \bibinfo{author}{Dhayal, S.},
  \bibinfo{author}{Mathur, M.} \& \bibinfo{author}{Ramaniah, L.~M.}
\newblock \bibinfo{journal}{\bibinfo{title}{Structural and energetic properties
  of $\alpha$-amino acids: {A} first principles density functional study}}.
\newblock {\emph{\JournalTitle{Molecular Physics}}}
  \textbf{\bibinfo{volume}{106}}, \bibinfo{pages}{2289--2300}
  (\bibinfo{year}{2008}).

\bibitem{selvarengan2004potential}
\bibinfo{author}{Selvarengan, P.} \& \bibinfo{author}{Kolandaivel, P.}
\newblock \bibinfo{journal}{\bibinfo{title}{Potential energy surface study on
  glycine, alanine and their zwitterionic forms}}.
\newblock {\emph{\JournalTitle{Journal of Molecular Structure: THEOCHEM}}}
  \textbf{\bibinfo{volume}{671}}, \bibinfo{pages}{77--86}
  (\bibinfo{year}{2004}).

\bibitem{csaszar1999ab}
\bibinfo{author}{Cs{\'a}sz{\'a}r, A.~G.} \& \bibinfo{author}{Perczel, A.}
\newblock \bibinfo{journal}{\bibinfo{title}{Ab initio characterization of
  building units in peptides and proteins}}.
\newblock {\emph{\JournalTitle{Progress in Biophysics and Molecular Biology}}}
  \textbf{\bibinfo{volume}{71}}, \bibinfo{pages}{243--309}
  (\bibinfo{year}{1999}).

\bibitem{schlund2008conformational}
\bibinfo{author}{Schlund, S.}, \bibinfo{author}{M{\"u}ller, R.},
  \bibinfo{author}{Gra$\beta$mann, C.} \& \bibinfo{author}{Engels, B.}
\newblock \bibinfo{journal}{\bibinfo{title}{Conformational analysis of arginine
  in gas phase--{A} strategy for scanning the potential energy surface
  effectively}}.
\newblock {\emph{\JournalTitle{Journal of Computational Chemistry}}}
  \textbf{\bibinfo{volume}{29}}, \bibinfo{pages}{407--415}
  (\bibinfo{year}{2008}).

\bibitem{riffet2011acid}
\bibinfo{author}{Riffet, V.}, \bibinfo{author}{Frison, G.} \&
  \bibinfo{author}{Bouchoux, G.}
\newblock \bibinfo{journal}{\bibinfo{title}{Acid--base thermochemistry of
  gaseous oxygen and sulfur substituted amino acids ({S}er, {T}hr, {C}ys,
  {M}et)}}.
\newblock {\emph{\JournalTitle{Physical Chemistry Chemical Physics}}}
  \textbf{\bibinfo{volume}{13}}, \bibinfo{pages}{18561--18580}
  (\bibinfo{year}{2011}).

\bibitem{baek2011density}
\bibinfo{author}{Baek, K.}, \bibinfo{author}{Fujimura, Y.},
  \bibinfo{author}{Hayashi, M.}, \bibinfo{author}{Lin, S.} \&
  \bibinfo{author}{Kim, S.}
\newblock \bibinfo{journal}{\bibinfo{title}{Density functional theory study of
  conformation-dependent properties of neutral and radical cationic
  {L}-tyrosine and {L}-tryptophan}}.
\newblock {\emph{\JournalTitle{The Journal of Physical Chemistry A}}}
  \textbf{\bibinfo{volume}{115}}, \bibinfo{pages}{9658--9668}
  (\bibinfo{year}{2011}).

\bibitem{floris2012density}
\bibinfo{author}{Floris, F.~M.}, \bibinfo{author}{Filippi, C.} \&
  \bibinfo{author}{Amovilli, C.}
\newblock \bibinfo{journal}{\bibinfo{title}{A density functional and quantum
  {M}onte {C}arlo study of glutamic acid in vacuo and in a dielectric continuum
  medium}}.
\newblock {\emph{\JournalTitle{The Journal of Chemical Physics}}}
  \textbf{\bibinfo{volume}{137}}, \bibinfo{pages}{075102}
  (\bibinfo{year}{2012}).

\bibitem{ropo2016first}
\bibinfo{author}{Ropo, M.}, \bibinfo{author}{Schneider, M.},
  \bibinfo{author}{Baldauf, C.} \& \bibinfo{author}{Blum, V.}
\newblock \bibinfo{journal}{\bibinfo{title}{{First-principles data set of
  45,892 isolated and cation-coordinated conformers of 20 proteinogenic amino
  acids}}}.
\newblock {\emph{\JournalTitle{Scientific Data}}} \textbf{\bibinfo{volume}{3}},
  \bibinfo{pages}{1--13} (\bibinfo{year}{2016}).

\bibitem{huang1995biostructural}
\bibinfo{author}{Huang, H.}, \bibinfo{author}{Li, D.} \&
  \bibinfo{author}{Cowan, J.}
\newblock \bibinfo{journal}{\bibinfo{title}{Biostructural chemistry of
  magnesium. regulation of mithramycin-{DNA} interactions by {M}g$^{2+}$
  coordination}}.
\newblock {\emph{\JournalTitle{Biochimie}}} \textbf{\bibinfo{volume}{77}},
  \bibinfo{pages}{729--738} (\bibinfo{year}{1995}).

\bibitem{romani2011cellular}
\bibinfo{author}{Romani, A.~M.}
\newblock \bibinfo{journal}{\bibinfo{title}{Cellular magnesium homeostasis}}.
\newblock {\emph{\JournalTitle{Archives of biochemistry and biophysics}}}
  \textbf{\bibinfo{volume}{512}}, \bibinfo{pages}{1--23}
  (\bibinfo{year}{2011}).

\bibitem{forsen1994calcium}
\bibinfo{author}{Forsen, S.} \& \bibinfo{author}{Kordel, J.}
\newblock \bibinfo{title}{Calcium in biological systems}
  (\bibinfo{year}{1994}).

\bibitem{grauffel2019cellular}
\bibinfo{author}{Grauffel, C.}, \bibinfo{author}{Dudev, T.} \&
  \bibinfo{author}{Lim, C.}
\newblock \bibinfo{journal}{\bibinfo{title}{Why cellular di/triphosphates
  preferably bind {M}g$^{2+}$ and not {C}a$^{2+}$}}.
\newblock {\emph{\JournalTitle{Journal of Chemical Theory and Computation}}}
  \textbf{\bibinfo{volume}{15}}, \bibinfo{pages}{6992--7003}
  (\bibinfo{year}{2019}).

\bibitem{mahmoud2012functionalized}
\bibinfo{author}{Mahmoud, W.~E.}
\newblock \bibinfo{journal}{\bibinfo{title}{Functionalized {ME}-capped {C}d{S}e
  quantum dots based luminescence probe for detection of {B}a$^{2+}$ ions}}.
\newblock {\emph{\JournalTitle{Sensors and Actuators B: Chemical}}}
  \textbf{\bibinfo{volume}{164}}, \bibinfo{pages}{76--81}
  (\bibinfo{year}{2012}).

\bibitem{wilkinson2016fair}
\bibinfo{author}{Wilkinson, M.~D.} \emph{et~al.}
\newblock \bibinfo{journal}{\bibinfo{title}{{The FAIR Guiding Principles for
  scientific data management and stewardship}}}.
\newblock {\emph{\JournalTitle{Scientific Data}}} \textbf{\bibinfo{volume}{3}},
  \bibinfo{pages}{1--9} (\bibinfo{year}{2016}).

\bibitem{wittenburg2020fair}
\bibinfo{author}{Wittenburg, P.}, \bibinfo{author}{Lautenschlager, M.},
  \bibinfo{author}{Thiemann, H.}, \bibinfo{author}{Baldauf, C.} \&
  \bibinfo{author}{Trilsbeek, P.}
\newblock \bibinfo{journal}{\bibinfo{title}{{FAIR} practices in {E}urope}}.
\newblock {\emph{\JournalTitle{Data Intelligence}}}
  \textbf{\bibinfo{volume}{2}}, \bibinfo{pages}{257--263}
  (\bibinfo{year}{2020}).

\bibitem{noy2001ontology}
\bibinfo{author}{Noy, N.~F.}, \bibinfo{author}{McGuinness, D.~L.} \emph{et~al.}
\newblock \bibinfo{title}{{Ontology development 101: A guide to creating your
  first ontology}} (\bibinfo{year}{2001}).

\bibitem{wales1997global}
\bibinfo{author}{Wales, D.~J.} \& \bibinfo{author}{Doye, J.~P.}
\newblock \bibinfo{journal}{\bibinfo{title}{{Global optimization by
  basin-hopping and the lowest energy structures of Lennard-Jones clusters
  containing up to 110 atoms}}}.
\newblock {\emph{\JournalTitle{The Journal of Physical Chemistry A}}}
  \textbf{\bibinfo{volume}{101}}, \bibinfo{pages}{5111--5116}
  (\bibinfo{year}{1997}).

\bibitem{wales1999global}
\bibinfo{author}{Wales, D.~J.} \& \bibinfo{author}{Scheraga, H.~A.}
\newblock \bibinfo{journal}{\bibinfo{title}{{Global optimization of clusters,
  crystals, and biomolecules}}}.
\newblock {\emph{\JournalTitle{Science}}} \textbf{\bibinfo{volume}{285}},
  \bibinfo{pages}{1368--1372} (\bibinfo{year}{1999}).

\bibitem{jorgensen1996development}
\bibinfo{author}{Jorgensen, W.~L.}, \bibinfo{author}{Maxwell, D.~S.} \&
  \bibinfo{author}{Tirado-Rives, J.}
\newblock \bibinfo{journal}{\bibinfo{title}{{Development and testing of the
  OPLS all-atom force field on conformational energetics and properties of
  organic liquids}}}.
\newblock {\emph{\JournalTitle{Journal of the American Chemical Society}}}
  \textbf{\bibinfo{volume}{118}}, \bibinfo{pages}{11225--11236}
  (\bibinfo{year}{1996}).

\bibitem{blum2009ab}
\bibinfo{author}{Blum, V.} \emph{et~al.}
\newblock \bibinfo{journal}{\bibinfo{title}{{Ab initio molecular simulations
  with numeric atom-centered orbitals}}}.
\newblock {\emph{\JournalTitle{Computer Physics Communications}}}
  \textbf{\bibinfo{volume}{180}}, \bibinfo{pages}{2175--2196}
  (\bibinfo{year}{2009}).

\bibitem{havu2009efficient}
\bibinfo{author}{Havu, V.}, \bibinfo{author}{Blum, V.}, \bibinfo{author}{Havu,
  P.} \& \bibinfo{author}{Scheffler, M.}
\newblock \bibinfo{journal}{\bibinfo{title}{{Efficient O (N) integration for
  all-electron electronic structure calculation using numeric basis
  functions}}}.
\newblock {\emph{\JournalTitle{Journal of Computational Physics}}}
  \textbf{\bibinfo{volume}{228}}, \bibinfo{pages}{8367--8379}
  (\bibinfo{year}{2009}).

\bibitem{ren2012resolution}
\bibinfo{author}{Ren, X.} \emph{et~al.}
\newblock \bibinfo{journal}{\bibinfo{title}{{Resolution-of-identity approach to
  Hartree--Fock, hybrid density functionals, RPA, MP2 and GW with numeric
  atom-centered orbital basis functions}}}.
\newblock {\emph{\JournalTitle{New Journal of Physics}}}
  \textbf{\bibinfo{volume}{14}}, \bibinfo{pages}{053020}
  (\bibinfo{year}{2012}).

\bibitem{perdew1996generalized}
\bibinfo{author}{Perdew, J.~P.}, \bibinfo{author}{Burke, K.} \&
  \bibinfo{author}{Ernzerhof, M.}
\newblock \bibinfo{journal}{\bibinfo{title}{{Generalized gradient approximation
  made simple}}}.
\newblock {\emph{\JournalTitle{Physical Review Letters}}}
  \textbf{\bibinfo{volume}{77}}, \bibinfo{pages}{3865} (\bibinfo{year}{1996}).

\bibitem{tkatchenko2009accurate}
\bibinfo{author}{Tkatchenko, A.} \& \bibinfo{author}{Scheffler, M.}
\newblock \bibinfo{journal}{\bibinfo{title}{{Accurate molecular van der Waals
  interactions from ground-state electron density and free-atom reference
  data}}}.
\newblock {\emph{\JournalTitle{Physical Review Letters}}}
  \textbf{\bibinfo{volume}{102}}, \bibinfo{pages}{073005}
  (\bibinfo{year}{2009}).

\bibitem{swendsen1986replica}
\bibinfo{author}{Swendsen, R.~H.} \& \bibinfo{author}{Wang, J.-S.}
\newblock \bibinfo{journal}{\bibinfo{title}{{Replica Monte Carlo simulation of
  spin-glasses}}}.
\newblock {\emph{\JournalTitle{Physical Review Letters}}}
  \textbf{\bibinfo{volume}{57}}, \bibinfo{pages}{2607} (\bibinfo{year}{1986}).

\bibitem{sugita1999replica}
\bibinfo{author}{Sugita, Y.} \& \bibinfo{author}{Okamoto, Y.}
\newblock \bibinfo{journal}{\bibinfo{title}{{Replica-exchange molecular
  dynamics method for protein folding}}}.
\newblock {\emph{\JournalTitle{Chemical Physics Letters}}}
  \textbf{\bibinfo{volume}{314}}, \bibinfo{pages}{141--151}
  (\bibinfo{year}{1999}).

\bibitem{wong1979algorithm}
\bibinfo{author}{Wong, M.~A.} \& \bibinfo{author}{Hartigan, J.}
\newblock \bibinfo{journal}{\bibinfo{title}{{Algorithm as 136: A k-means
  clustering algorithm}}}.
\newblock {\emph{\JournalTitle{Journal of the Royal Statistical Society. Series
  C (Applied Statistics)}}} \textbf{\bibinfo{volume}{28}},
  \bibinfo{pages}{100--108} (\bibinfo{year}{1979}).

\bibitem{hirshfeld1977bonded}
\bibinfo{author}{Hirshfeld, F.~L.}
\newblock \bibinfo{journal}{\bibinfo{title}{{Bonded-atom fragments for
  describing molecular charge densities}}}.
\newblock {\emph{\JournalTitle{Theoretica Chimica Acta}}}
  \textbf{\bibinfo{volume}{44}}, \bibinfo{pages}{129--138}
  (\bibinfo{year}{1977}).

\bibitem{distasio2014many}
\bibinfo{author}{DiStasio, R.~A.}, \bibinfo{author}{Gobre, V.~V.} \&
  \bibinfo{author}{Tkatchenko, A.}
\newblock \bibinfo{journal}{\bibinfo{title}{{Many-body van der Waals
  interactions in molecules and condensed matter}}}.
\newblock {\emph{\JournalTitle{Journal of Physics: Condensed Matter}}}
  \textbf{\bibinfo{volume}{26}}, \bibinfo{pages}{213202}
  (\bibinfo{year}{2014}).

\bibitem{henkelman2006fast}
\bibinfo{author}{Henkelman, G.}, \bibinfo{author}{Arnaldsson, A.} \&
  \bibinfo{author}{J{\'o}nsson, H.}
\newblock \bibinfo{journal}{\bibinfo{title}{{A fast and robust algorithm for
  Bader decomposition of charge density}}}.
\newblock {\emph{\JournalTitle{Computational Materials Science}}}
  \textbf{\bibinfo{volume}{36}}, \bibinfo{pages}{354--360}
  (\bibinfo{year}{2006}).

\bibitem{sanville2007improved}
\bibinfo{author}{Sanville, E.}, \bibinfo{author}{Kenny, S.~D.},
  \bibinfo{author}{Smith, R.} \& \bibinfo{author}{Henkelman, G.}
\newblock \bibinfo{journal}{\bibinfo{title}{{Improved grid-based algorithm for
  Bader charge allocation}}}.
\newblock {\emph{\JournalTitle{Journal of Computational Chemistry}}}
  \textbf{\bibinfo{volume}{28}}, \bibinfo{pages}{899--908}
  (\bibinfo{year}{2007}).

\bibitem{yu2011accurate}
\bibinfo{author}{Yu, M.} \& \bibinfo{author}{Trinkle, D.~R.}
\newblock \bibinfo{journal}{\bibinfo{title}{{Accurate and efficient algorithm
  for Bader charge integration}}}.
\newblock {\emph{\JournalTitle{The Journal of Chemical Physics}}}
  \textbf{\bibinfo{volume}{134}}, \bibinfo{pages}{064111}
  (\bibinfo{year}{2011}).

\bibitem{bayly1993well}
\bibinfo{author}{Bayly, C.~I.}, \bibinfo{author}{Cieplak, P.},
  \bibinfo{author}{Cornell, W.} \& \bibinfo{author}{Kollman, P.~A.}
\newblock \bibinfo{journal}{\bibinfo{title}{{A well-behaved electrostatic
  potential based method using charge restraints for deriving atomic charges:
  the RESP model}}}.
\newblock {\emph{\JournalTitle{The Journal of Physical Chemistry}}}
  \textbf{\bibinfo{volume}{97}}, \bibinfo{pages}{10269--10280}
  (\bibinfo{year}{1993}).

\bibitem{singh1984approach}
\bibinfo{author}{Singh, U.~C.} \& \bibinfo{author}{Kollman, P.~A.}
\newblock \bibinfo{journal}{\bibinfo{title}{{An approach to computing
  electrostatic charges for molecules}}}.
\newblock {\emph{\JournalTitle{Journal of Computational Chemistry}}}
  \textbf{\bibinfo{volume}{5}}, \bibinfo{pages}{129--145}
  (\bibinfo{year}{1984}).

\bibitem{fox1998application}
\bibinfo{author}{Fox, T.} \& \bibinfo{author}{Kollman, P.~A.}
\newblock \bibinfo{journal}{\bibinfo{title}{{Application of the RESP
  methodology in the parametrization of organic solvents}}}.
\newblock {\emph{\JournalTitle{The Journal of Physical Chemistry B}}}
  \textbf{\bibinfo{volume}{102}}, \bibinfo{pages}{8070--8079}
  (\bibinfo{year}{1998}).

\bibitem{wang2001antechamber}
\bibinfo{author}{Wang, J.}, \bibinfo{author}{Wang, W.},
  \bibinfo{author}{Kollman, P.~A.} \& \bibinfo{author}{Case, D.~A.}
\newblock \bibinfo{journal}{\bibinfo{title}{{Antechamber: an accessory software
  package for molecular mechanical calculations}}}.
\newblock {\emph{\JournalTitle{J. Am. Chem. Soc}}}
  \textbf{\bibinfo{volume}{222}}, \bibinfo{pages}{U403} (\bibinfo{year}{2001}).

\bibitem{salomon2013overview}
\bibinfo{author}{Salomon-Ferrer, R.}, \bibinfo{author}{Case, D.~A.} \&
  \bibinfo{author}{Walker, R.~C.}
\newblock \bibinfo{journal}{\bibinfo{title}{{An overview of the Amber
  biomolecular simulation package}}}.
\newblock {\emph{\JournalTitle{Wiley Interdisciplinary Reviews: Computational
  Molecular Science}}} \textbf{\bibinfo{volume}{3}}, \bibinfo{pages}{198--210}
  (\bibinfo{year}{2013}).

\bibitem{o2011open}
\bibinfo{author}{O'Boyle, N.~M.} \emph{et~al.}
\newblock \bibinfo{journal}{\bibinfo{title}{{Open Babel: An open Chemical
  toolbox}}}.
\newblock {\emph{\JournalTitle{Journal of Cheminformatics}}}
  \textbf{\bibinfo{volume}{3}}, \bibinfo{pages}{1--14} (\bibinfo{year}{2011}).

\bibitem{jo2008charmm}
\bibinfo{author}{Jo, S.}, \bibinfo{author}{Kim, T.}, \bibinfo{author}{Iyer,
  V.~G.} \& \bibinfo{author}{Im, W.}
\newblock \bibinfo{journal}{\bibinfo{title}{{CHARMM-GUI: a web-based graphical
  user interface for CHARMM}}}.
\newblock {\emph{\JournalTitle{Journal of Computational Chemistry}}}
  \textbf{\bibinfo{volume}{29}}, \bibinfo{pages}{1859--1865}
  (\bibinfo{year}{2008}).

\bibitem{eastman2017openmm}
\bibinfo{author}{Eastman, P.} \emph{et~al.}
\newblock \bibinfo{journal}{\bibinfo{title}{{OpenMM 7: Rapid development of
  high performance algorithms for molecular dynamics}}}.
\newblock {\emph{\JournalTitle{PLoS Computational Biology}}}
  \textbf{\bibinfo{volume}{13}}, \bibinfo{pages}{e1005659}
  (\bibinfo{year}{2017}).

\bibitem{Hu2021cation}
\bibinfo{author}{Hu, X.} \& \bibinfo{author}{Baldauf, C.}
\newblock \bibinfo{journal}{\bibinfo{title}{{Cation-coordinated conformers of
  20 proteinogenic amino acids with different protonation states}}}.
\newblock {\emph{\JournalTitle{NOMAD}}}
  \url{https://dx.doi.org/10.17172/NOMAD/2021.02.10-1} (\bibinfo{year}{2021}).

\bibitem{draxl2019nomad}
\bibinfo{author}{Draxl, C.} \& \bibinfo{author}{Scheffler, M.}
\newblock \bibinfo{journal}{\bibinfo{title}{{The NOMAD laboratory: from data
  sharing to artificial intelligence}}}.
\newblock {\emph{\JournalTitle{Journal of Physics: Materials}}}
  \textbf{\bibinfo{volume}{2}}, \bibinfo{pages}{036001} (\bibinfo{year}{2019}).

\bibitem{Hu2021ontology}
\bibinfo{author}{Hu, X.}, \bibinfo{author}{Lenz-Himmer, M.~O.} \&
  \bibinfo{author}{Baldauf, C.}
\newblock \bibinfo{journal}{\bibinfo{title}{{The ontology representation for a
  data set of cation-coordinated conformers of 20 proteinogenic amino acids
  with different protonation states}}}.
\newblock {\emph{\JournalTitle{EDMOND}}} \url{https://dx.doi.org/10.17617/3.5q}
  (\bibinfo{year}{2021}).

\bibitem{al2020automatic}
\bibinfo{author}{Al-Aswadi, F.~N.}, \bibinfo{author}{Chan, H.~Y.} \&
  \bibinfo{author}{Gan, K.~H.}
\newblock \bibinfo{journal}{\bibinfo{title}{{Automatic ontology construction
  from text: a review from shallow to deep learning trend}}}.
\newblock {\emph{\JournalTitle{Artificial Intelligence Review}}}
  \textbf{\bibinfo{volume}{53}}, \bibinfo{pages}{3901--3928}
  (\bibinfo{year}{2020}).

\bibitem{musen2015protege}
\bibinfo{author}{Musen, M.~A.}
\newblock \bibinfo{journal}{\bibinfo{title}{{The prot{\'e}g{\'e} project: a
  look back and a look forward}}}.
\newblock {\emph{\JournalTitle{AI Matters}}} \textbf{\bibinfo{volume}{1}},
  \bibinfo{pages}{4--12} (\bibinfo{year}{2015}).

\bibitem{lamy2017owlready}
\bibinfo{author}{Lamy, J.-B.}
\newblock \bibinfo{journal}{\bibinfo{title}{{Owlready: Ontology-oriented
  programming in Python with automatic classification and high level constructs
  for biomedical ontologies}}}.
\newblock {\emph{\JournalTitle{Artificial intelligence in medicine}}}
  \textbf{\bibinfo{volume}{80}}, \bibinfo{pages}{11--28}
  (\bibinfo{year}{2017}).

\bibitem{tsarkov2006fact++}
\bibinfo{author}{Tsarkov, D.} \& \bibinfo{author}{Horrocks, I.}
\newblock \bibinfo{title}{{FaCT++ description logic reasoner: System
  description}}.
\newblock In \emph{\bibinfo{booktitle}{International Joint Conference on
  Automated Reasoning}}, \bibinfo{pages}{292--297}
  (\bibinfo{organization}{Springer}, \bibinfo{year}{2006}).

\bibitem{wang2012development}
\bibinfo{author}{Wang, J.} \emph{et~al.}
\newblock \bibinfo{journal}{\bibinfo{title}{{Development of polarizable models
  for molecular mechanical calculations. 4. van der Waals parametrization}}}.
\newblock {\emph{\JournalTitle{The Journal of Physical Chemistry B}}}
  \textbf{\bibinfo{volume}{116}}, \bibinfo{pages}{7088--7101}
  (\bibinfo{year}{2012}).

\bibitem{li2017machine}
\bibinfo{author}{Li, Y.} \emph{et~al.}
\newblock \bibinfo{journal}{\bibinfo{title}{{Machine learning force field
  parameters from ab initio data}}}.
\newblock {\emph{\JournalTitle{Journal of Chemical Theory and Computation}}}
  \textbf{\bibinfo{volume}{13}}, \bibinfo{pages}{4492--4503}
  (\bibinfo{year}{2017}).

\bibitem{cole2016biomolecular}
\bibinfo{author}{Cole, D.~J.}, \bibinfo{author}{Vilseck, J.~Z.},
  \bibinfo{author}{Tirado-Rives, J.}, \bibinfo{author}{Payne, M.~C.} \&
  \bibinfo{author}{Jorgensen, W.~L.}
\newblock \bibinfo{journal}{\bibinfo{title}{{Biomolecular force field
  parameterization via atoms-in-molecule electron density partitioning}}}.
\newblock {\emph{\JournalTitle{Journal of Chemical Theory and Computation}}}
  \textbf{\bibinfo{volume}{12}}, \bibinfo{pages}{2312--2323}
  (\bibinfo{year}{2016}).

\bibitem{rai2013fast}
\bibinfo{author}{Rai, B.~K.} \& \bibinfo{author}{Bakken, G.~A.}
\newblock \bibinfo{journal}{\bibinfo{title}{{Fast and accurate generation of ab
  initio quality atomic charges using nonparametric statistical regression}}}.
\newblock {\emph{\JournalTitle{Journal of Computational Chemistry}}}
  \textbf{\bibinfo{volume}{34}}, \bibinfo{pages}{1661--1671}
  (\bibinfo{year}{2013}).

\bibitem{bleiziffer2018machine}
\bibinfo{author}{Bleiziffer, P.}, \bibinfo{author}{Schaller, K.} \&
  \bibinfo{author}{Riniker, S.}
\newblock \bibinfo{journal}{\bibinfo{title}{{Machine learning of partial
  charges derived from high-quality quantum-mechanical calculations}}}.
\newblock {\emph{\JournalTitle{Journal of Chemical Information and Modeling}}}
  \textbf{\bibinfo{volume}{58}}, \bibinfo{pages}{579--590}
  (\bibinfo{year}{2018}).

\bibitem{moller1934note}
\bibinfo{author}{M{\o}ller, C.} \& \bibinfo{author}{Plesset, M.~S.}
\newblock \bibinfo{journal}{\bibinfo{title}{{Note on an approximation treatment
  for many-electron systems}}}.
\newblock {\emph{\JournalTitle{Physical Review}}}
  \textbf{\bibinfo{volume}{46}}, \bibinfo{pages}{618} (\bibinfo{year}{1934}).

\bibitem{head1988mp2}
\bibinfo{author}{Head-Gordon, M.}, \bibinfo{author}{Pople, J.~A.} \&
  \bibinfo{author}{Frisch, M.~J.}
\newblock \bibinfo{journal}{\bibinfo{title}{{MP2 energy evaluation by direct
  methods}}}.
\newblock {\emph{\JournalTitle{Chemical Physics Letters}}}
  \textbf{\bibinfo{volume}{153}}, \bibinfo{pages}{503--506}
  (\bibinfo{year}{1988}).

\bibitem{ambrosetti2014long}
\bibinfo{author}{Ambrosetti, A.}, \bibinfo{author}{Reilly, A.~M.},
  \bibinfo{author}{DiStasio~Jr, R.~A.} \& \bibinfo{author}{Tkatchenko, A.}
\newblock \bibinfo{journal}{\bibinfo{title}{{Long-range correlation energy
  calculated from coupled atomic response functions}}}.
\newblock {\emph{\JournalTitle{The Journal of Chemical Physics}}}
  \textbf{\bibinfo{volume}{140}}, \bibinfo{pages}{18A508}
  (\bibinfo{year}{2014}).

\bibitem{riplinger2013efficient}
\bibinfo{author}{Riplinger, C.} \& \bibinfo{author}{Neese, F.}
\newblock \bibinfo{journal}{\bibinfo{title}{{An efficient and near linear
  scaling pair natural orbital based local coupled cluster method}}}.
\newblock {\emph{\JournalTitle{The Journal of Chemical Physics}}}
  \textbf{\bibinfo{volume}{138}}, \bibinfo{pages}{034106}
  (\bibinfo{year}{2013}).

\bibitem{riplinger2013natural}
\bibinfo{author}{Riplinger, C.}, \bibinfo{author}{Sandhoefer, B.},
  \bibinfo{author}{Hansen, A.} \& \bibinfo{author}{Neese, F.}
\newblock \bibinfo{journal}{\bibinfo{title}{{Natural triple excitations in
  local coupled cluster calculations with pair natural orbitals}}}.
\newblock {\emph{\JournalTitle{The Journal of Chemical Physics}}}
  \textbf{\bibinfo{volume}{139}}, \bibinfo{pages}{134101}
  (\bibinfo{year}{2013}).

\bibitem{supady2015first}
\bibinfo{author}{Supady, A.}, \bibinfo{author}{Blum, V.} \&
  \bibinfo{author}{Baldauf, C.}
\newblock \bibinfo{journal}{\bibinfo{title}{{First-principles molecular
  structure search with a genetic algorithm}}}.
\newblock {\emph{\JournalTitle{Journal of Chemical Information and Modeling}}}
  \textbf{\bibinfo{volume}{55}}, \bibinfo{pages}{2338--2348}
  (\bibinfo{year}{2015}).

\end{thebibliography}


\newpage
\section*{Acknowledgements} 


X.H. is grateful for a doctoral fellowship by the China Scholarship Council. M.O.L.H. and C.B. acknowledge funding by the Federal Ministry of Education and Research of Germany for the project STREAM (``Semantische Repräsentation, Vernetzung und Kuratierung von qualitätsgesicherten Materialdaten'', ID: 16QK11C).

\section*{Author contributions statement}


X.H. performed the calculations of all conformers, curated the data, constructed the ontology, and wrote the manuscript. M.L. helped with the construction of ontology and contributed to the manuscript. C.B. designed the study, curated the data, and wrote the manuscript.

\section*{Competing interests} 


The authors declare no competing financial interests.

\clearpage

\section*{Figures \& Tables}

\newpage

\newpage

\newpage

\newpage

\newpage

\newpage

\clearpage

\begin{table}
\caption{\label{tab:filetypes}List and description of file types in the data set. }
\begin{tabulary}{0.95\linewidth}{lLc}
\hline
\textbf{File name} & \textbf{Description}  & \textbf{Code/Format}\\ \hline
\multicolumn{3}{c}{\textbf{FHI-aims Input Files}} \\  \hline
\texttt{geometry.in} & Cartesian coordinates of the complexes & FHI-aims \\
\texttt{control.in} & Input file with technical parameters for electronic structure calculations & FHI-aims \\

\multicolumn{3}{c}{\textbf{FHI-aims Output Files}} \\  \hline
\texttt{FHIaims.out} & Main output the electronic structure calculations, contains: total energy, vdW energy and effective atomic volume etc. & FHI-aims \\
\multicolumn{2}{l}{\texttt{cube$_{-}$001$_{-}$total$_{-}$density.cube.bz2}} &\\
 & Cube file representation of the electron density (bzip2 compressed) & FHI-aims \\
\multicolumn{2}{l}{\texttt{potential$_{-}$esp$_{-}$1.cube.bz2}} &\\
 & Cube file representation of the electrostatic potential (bzip2 compressed) & FHI-aims \\
 \texttt{hirsh.chrg} & Hirshfeld charges & self-made \\

\multicolumn{3}{c}{\textbf{Geometries}} \\ \hline
\texttt{coords.xyz} & coordinate file & \textit{xyz} format  \\
\texttt{residue.pdb} & coordinate file & CHARMM \\
\texttt{[cation].pdb} & separate coordinate file for each of the cations \texttt{Ca}, \texttt{Ba}, \texttt{Mg}& CHARMM  \\
\texttt{openmm.pdb} & coordinate file & openMM\\

\multicolumn{3}{c}{\textbf{Bader AIM calculations}} \\ \hline
\multicolumn{2}{l}{\texttt{ACF.dat, AVF.dat, BCF.dat, bader.out}} &\\
 & information of Bader charge analysis & Bader \\
\multicolumn{2}{l}{\texttt{nuclear.chrg, bdr.chrg}} &\\
 & information of Bader charge analysis & self-made \\
\texttt{bader.chrg} & Bader charges & self-made \\

\multicolumn{3}{c}{\textbf{RESP calculations}} \\ \hline
\multicolumn{2}{l}{\texttt{geometry.respout2, resp$_{-}$files.tar.gz}} &\\
 & RESP charge information & Antechamber \\
\texttt{resp.chrg} & RESP charges &self-made \\

\multicolumn{3}{c}{\textbf{Aggregated output}} \\  \hline
\texttt{geometry.ext} & collection of coordinate and charge information & self-made \\
\texttt{energies.dat} & collection of total energy and interaction energy & self-made \\
    \hline
\end{tabulary}
\end{table}

\clearpage
\end{document}